\documentclass[final,5p,twocolumn]{elsarticle}
\usepackage[colorlinks=true, allcolors=blue]{hyperref} 
\usepackage{amssymb}
\usepackage{amsmath}
\usepackage{enumitem} 
\graphicspath{{Figures/}}
\usepackage[caption=true]{subfig}
\usepackage{xfrac}
\usepackage{multirow}
\usepackage{cleveref}
\usepackage{bbm}
\usepackage{bm}
\usepackage{balance}
\usepackage{array}
\usepackage{booktabs}
\usepackage{float}
\usepackage[linesnumbered, ruled, vlined]{algorithm2e}
\usepackage{datatool} 
\DTLsetseparator{,}
\DTLloaddb{theta}{Data/Parameters.txt}
\DTLloaddb{m}{Data/Moments.txt}
\usepackage[figuresleft]{rotating}

\usepackage{tikz}
\usetikzlibrary{shapes.geometric, arrows}
\usepackage{amsthm}
\theoremstyle{definition}

\newtheorem{remark}{Remark}

\newcommand{\bc}{\begin{center}}
\newcommand{\ec}{\end{center}}
\newcommand{\be}{\begin{equation}}
\newcommand{\ee}{\end{equation}}
\newcommand{\bea}{\begin{eqnarray}}
\newcommand{\eea}{\end{eqnarray}}
\newcommand{\beq}{\begin{eqnarray*}}
\newcommand{\eeq}{\end{eqnarray*}}
\newcommand{\bv}{\left( \begin{array}{c} }
\newcommand{\ev}{\end{array} \right) }

\newcommand{\Var}{\mbox{Var}}
\newcommand{\Cov}{\mbox{Cov}}

\newcommand{\diag}{\mathrm{diag}}

\newcommand{\ve}[1]{\bm{#1}}

\makeatletter
\def\ps@pprintTitle{%
  \let\@oddhead\@empty
  \let\@evenhead\@empty
  \def\@oddfoot{\reset@font\hfil\thepage\hfil}
  \let\@evenfoot\@oddfoot
}
\makeatother

\begin{document}
\begin{filecontents*}{Data/Parameters.txt}
Variable,Lower,Estimate,Upper
Nh,3.609,5.992,8.376
Delta,0.058,0.092,0.125
Kappa,1.716,2.555,3.394
Nu,2.876,3.381,3.887
Sigma,0.08,0.418,0.756
\end{filecontents*}

\begin{filecontents*}{Data/Moments.txt}
Moment,Lower,Estimate,Upper,Empirical
Mean,-0.0,-0.0,0.0,0.0
StandardDeviation,0.0,0.0,0.0,0.001
Kurtosis,-1195.441,2323.204,5841.849,4.12
KolmogorovSmirnov,0.424,0.456,0.488,0.0
Hurst,0.488,0.527,0.566,0.231
GPH,-0.199,0.055,0.309,0.729
ADF,-311.74,-296.876,-282.012,-176.529
GARCH,0.859,1.033,1.207,1.005
Hill,0.676,0.958,1.241,2.043
\end{filecontents*}

\begin{frontmatter}
    \title{Simulation and estimation of an agent-based market-model with a matching engine}
    
    \author[a1]{Ivan Jericevich}
    \ead{jrciva001@myuct.ac.za}
    \author[a1,a2]{Patrick Chang}
    \ead{patrick.chang@eng.ox.ac.uk}
    \author[a1]{Tim Gebbie}
    \ead{tim.gebbie@uct.ac.za}
    
    \address[a1]{Department of Statistical Sciences, University of Cape Town, Rondebosch 7700, South Africa} 
    \address[a2]{Department of Engineering Science, University of Oxford, Oxford OX1 3PJ, United Kingdom}
    
    \begin{abstract}
    An agent-based model with interacting low frequency liquidity takers inter-mediated by high-frequency liquidity providers acting collectively as market makers can be used to provide realistic simulated price impact curves. This is possible when agent-based model interactions occur asynchronously via order matching using a matching engine in event time to replace sequential calendar time market clearing. Here the matching engine infrastructure has been modified to provide a continuous feed of order confirmations and updates as message streams in order to conform more closely to live trading environments. The resulting trade and quote message data from the simulations are then aggregated, calibrated and visualised. Various stylised facts are presented along with event visualisations and price impact curves. We argue that additional realism in modelling can be achieved with a small set of agent parameters and simple interaction rules once interactions are reactive, asynchronous and in event time. We argue that the reactive nature of market agents may be a fundamental property of financial markets and when accounted for can allow for parsimonious modelling without recourse to additional sources of noise.
    \end{abstract}
    \begin{keyword}
        market matching-engine \sep agent-based model \sep simulation \sep estimation \sep calibration \sep price-impact
    \end{keyword}
\end{frontmatter}

\tableofcontents


\section{Introduction \label{sec:Introduction}}

The pairing of agent-based models with a matching engine in an artificial stock market framework that mimics the real-world work-flow and system implementation offers a controlled environment for exploring market complexity without forcing the market structure and constraints into the domain specification of agent interaction rules.

Financial Agent-Based Models (ABM's) are characterised by a bottom-up approach to modelling complex systems by attributing market dynamics and emergent phenomena to relatively simple interactions of trader agents without the need to explicitly program or mechanistically define a vast web of complex generative models to capture all the nuances of real financial markets. These models are appealing for, among other things, being able to produce reasonably realistic financial time series that can be used to inform market regulation decisions, explain market phenomena, and to provide test-beds for testing trading strategies in an artificial exchange. 

ABM's successfully link the micro-level rules of investor interactions with the macro-behaviour of asset prices in real markets. However, not all complex behaviour in markets can be replicated using this bottom-up approach. There appears to be a hierarchy in the levels of abstractions which lead to different nuances in overall behaviour \cite{plattgebbie2016problem,WilcoxGebbie2015} when observed from different averaging scales and from different market participants. For example, the concerns of an equity risk trader are not the same as those of a commodity trader, or a mutual fund manager or a fixed income manager, or that of a central bank market regulator and economic policy makers. Nonetheless, ABM's are interesting as a tool for replicating complex behaviour.\footnote{For extensive reviews of heterogeneous agent models and computational agent models refer to \cite{bonabeau2002agent,chakraborti2011econophysics2,chiarella2009heterogeneity,dieci2018heterogeneous,hommes2006heterogeneous,lebaron2000agent, lebaron2006agent,wang2018agent}.}

Here we replicate real equity markets at the level of the model components’ design as well as at the model’s output level. The recovery of the market orders' price impact being a crucial component \cite{farmer2005predictive,Alexandru2015}. \citet{farmer2005predictive} argued that with the zero-intelligence approach one may be able to find macroscopic averaged laws, not dissimilar to those used in thermodynamics, that can capture universal relationships between various macroscopic averaged quantities derived from microscopic interactions, such as the average diffusion rates, spreads and so on. Although a zero-intelligence approach is desirable, there is some motivation for considering perfectly rational or at least boundedly rational interacting agents, as in this case. Here microscopic rules are directly related to macroscopic properties via simulating a large number of agent interactions. The motivation is that some empirical features may not be straightforward to achieve without agent intelligence or strategic behaviour, or at least without detailed external assumptions about empirical facts and the distributional assumptions that may be needed to encode the macroscopic results from strategic agent interactions. 

In the financial ABM literature, significant attention has been devoted to the calibration of models involving closed-form solutions where markets are cleared at each iteration and prices are determined by a weighted average of trader expectations, {\it i.e.,} models with sequential market clearing which are analogous to closing auction trading sessions at the end of each day. This simplification is appropriate for mimicking the volume maximising mechanism typically used by an exchange. In contrast, intraday models such as those of \cite{leal2016rock,preis2006multi} focus on recreating the double-auction dynamics, but with prices that are still formed sequentially in calendar time. We deviate here by introducing a framework for semi-asynchronous order submission and matching by pairing an ABM with a matching engine. We consider this an important step towards embracing the event time realities of financial markets at the level of the model design.

At the present, there are several agent-based artificial exchanges, with varying functionalities and architectures, addressing different problems \cite{brandouy2013design, kumar2013implementing, lebaron2002building, refinitivwhitepaper}. Most of the present artificial market platforms suffer from a lack of flexibility and must be viewed as software frameworks and tool libraries rather than Advanced Programming Interfaces (API's). This is because they are mainly implemented for solving a specific problem confined to a particular choice of programming tools, and (most of the time) cannot easily be used to explore a wide range of financial issues that more closely mimic how software systems are designed and used for real trading and investment management \cite{brandouy2013design}. Most similar to our implementation, \citet{brandouy2013design} present the \href{https://github.com/cristal-smac/atom}{ArTificial Open Market} Java API (ATOM) through which one can build a variety of experiments on order-driven markets and discuss various design and architecture issues related to artificial exchanges. The difference here is that the implementation in this paper focuses on building agent-based models around well defined market mechanics. We first focus on defining the market rules and mechanics, and then allowing interaction via the rules and market mechanics. We believe that our implementation of separating the Model Management System (MMS) from the order matching system, the Matching Engine (ME), offers the desired flexibility for exploring different issues in financial markets without constraining the nature of the agent-based model (see Figure \ref{fig:flowchart}). This is because the order matching is then entirely decoupled from what agents do, and agent interactions do not take place in some globally synchronised calendar time.\footnote{A discussion on the nature of time in the trading environment can be found in \cite{CHANG2021126329}.}

This paper uses the CoinTossX matching engine software \cite{arxiv2021hawkes, arxiv2021cointossx, sing2017cointossx, sing2017jse} for the purpose of data generation and agent-based modelling in an artificial exchange. This software has now been extensively tested using simple point-processes \cite{arxiv2021cointossx,sing2017cointossx} as well as more complex multi-order type self-exciting point process test cases \cite{arxiv2021hawkes}. CoinTossX is interesting for, among other things, being able to closely replicate the mechanics of a real commercial exchanges which include different trading sessions (closing, opening, intraday and volatility auctions), different order-type and time-in-force combinations, and multiple securities and clients. It does all of this whilst maintaining low-latency, high-throughput and modularity of its components \cite{arxiv2021cointossx}. The matching engine places no constraints on the choice of time used by the agent-based model nor how the agents wish to interact with the market place --- merely that order messages be sent to the matching engine.

\subsection{Towards actor/event-based reactive systems}

The foundation of actor-based reactive systems is a set of individual components called actors or processes. These run autonomously in parallel and explicitly interact according to their local behaviour and reaction rules. The main fundamental differences between ABMs and reactive systems are \cite{crafa2021agent}: (i) a different management of time, and (ii) a fully decentralised decision control-logic. Both have significant effects on simulation dynamics. An agent-based model is typically iterated over stochastic or deterministic calendar time steps with aspects of the agents updated at each time step, whereas actors autonomously run in parallel and occasionally interact by exchanging messages. The sending of a message can be thought of as a signal that an event has occurred so that the receiver can react by modifying its behaviour and responding with another message \cite{hewitt2011actormodel,HBS1973actormodel}. Actors communicate asynchronously and may be ``blocked'' from further action until a reply or triggering event occurs. This asynchronicity is particularly well suited for financial markets and it can have significant effects on the model output. For agent-based models we argue that this may bring about a greater degree of realism, heterogeneity and computational efficiency. It also has the advantage of allowing modellers to simplify the models that specify how agents interact, {\it i.e.,} allow for models with fewer parameters that are still able to achieve high levels of realism. This may be also be an important step towards realising hierarchical causality within simulations of financial market systems \cite{WilcoxGebbie2015} so that both top-down and bottom-up sources of causality can be included within the agent-based modelling design.

\subsection{Intraday agent-based modelling}

While building our model, our thinking was informed by a number of different traditional intraday ABM approaches. We had to adapt them for an asynchronous continuous double auction implementation because of our departure from sequential market clearing.

In the \citet{leal2016rock} model, low frequency agents adopt trading rules based on chronological or calendar time, and can switch between fundamentalist and chartist strategies. High Frequency (HF) trader activation is event-driven, in the sense of a stochastic time, and depends on price fluctuations. These traders use directional strategies to exploit market information produced by Low-Frequency (LF) traders. The \citet{leal2016rock} model is our reference model and provides the basis for the ABM formulation in this paper --- not only because it is a simple model with relatively realistic properties, but because we have a well understood model calibration that is useful for our use case \cite{plattgebbie2016problem}. However, this model presents a number of challenges for integration with a matching engine. It uses sequential market clearing and thus it is not an asynchronous market. Second, the limit price placements follows more closely to a random walk setup. On the other hand, this is a useful specification because the model does not require a single market maker providing liquidity, rather the numerous high-frequency agents act as a proxy to achieve a similar effect. A general problem with market making that we carefully avoid is the need to repeatedly optimise and solve control problems to find optimal order placements; solving these type of problems generally only give placement for one level of the book.

With the same model, \citet{refinitivwhitepaper} develop an artificial exchange for data generation at scale. This heterogeneous model setup takes a traditional agent-based approach by specifying the same three agent types and parameters as provided by \citet{leal2016rock}. However, as an alternative to this, their software provides a library of models addressing specific use cases to produce realistic transaction price dynamics and trading scenarios. The four models in this library are: a heterogeneous model (the \citet{leal2016rock} model), a semi synthetic model, an asset interaction model and a bond pricing model. In these cases the synthetic market matches orders posted, publishing the bid, ask, execution prices and the volumes at all levels of the Limit-Order Book (LOB) at consecutive (but synchronous) time steps. Agents in the semi-synthetic model can either be ``real'' which is the case where order time, side and quantity are the same as that of historical data but the price is relative to prevailing market prices, or ``synthetic'' where the orders are completely defined by stochastic decision rules. The asset interaction model considers trading in multiple assets to induce correlated price dynamics where order prices follow a random walk and the sides and quantities are determined by a weighted combination of expected returns. Our approach is similar to that of the heterogeneous model but with asynchronous order submission times in a continuous double auction. In constructing an asynchronous continuous double auction ABM we draw from a number of noteworthy sources.

\citet{raman2019financial} apply deep learning methods to determine the directional bet. However, their limit order placement remains similar to a random walk. Their model allows for volume sizes to also partially depend on the strength of the directional signal predicted by the deep learning method. In this case the issue of implementation is the lack of a continuous double auction. However, some care is prudent when using machine learning methods in the presence of strategic agents as feature selection can induce feed-backs that change the target function(s).

\citet{preis2006multi} provide a limit order placement strategy that retains some of the ideas from \citet{farmer2005predictive} and also encapsulates the feedback dynamics from the market into the agents decisions. The calibration of this model has been explored in the literature \cite{plattgebbie2018can}. We draw inspiration from their ideas to the limit order placement for our HF agents. In their model, liquidity-provider-agents submit limit orders, and liquidity-taker-agents submit market orders only, but to the same LOB through a series of Monte-Carlo steps. During each Monte-Carlo step liquidity providers first submit limit orders based on the current state of the LOB, after which liquidity takers submit market orders. During each Monte-Carlo step liquidity-providers may also cancel previously placed orders based on uniform sampling and a probability parameter. Limit prices are set to be a certain amount of ticks away from the best on the contra side, the size of the tick is dependent on state variables and comes from a random variable. This point ensures that our HF agents will place LO's at different levels in the LOB while incorporating feedback dynamics. The additional modification required is that we need to have different volumes sizes.

\citet{Alexandru2015} is one of the few ABM setups that computes the resulting market impact arising from trading and order-book activity. Their setup requires some optimisation for decision making but is different from traditional optimal control and focuses on the appropriate market behaviour based on states of the LOB. Their work highlights the importance of agent decisions needing to depend on the state of the LOB to be realistic. In summary: The size of market orders should have some dependency on the spread, order-imbalance and depth of the LOB, and an overall supply-demand imbalance drives traders to price their orders more aggressively when their side of the book is crowded in order to increase their order execution probability (competition effect). Conversely, traders become less aggressive when the opposite side is thicker, forecasting a favourable short-term order flow (strategic effect).

The question that remains is how do we determine volume size? \citet{Alexandru2015} uses an independent log-normal distribution for the volume size. If we do something similar then we need to provide additional rules or constraints to ensure liquidity takers can remove sufficient liquidity because the current \citet{leal2016rock} model will not remove liquidity fast enough to maintain stability. Here we strive to avoid additional trading rules and constraints to force empirical realism, {\it e.g.,} by forcing HF agents to have entry thresholds as an additional parameter to be tuned to force empirical realism.

\subsection{Market making}

One key characteristic of the simulation environment that makes decision rules like order price setting challenging is that the market does not sequentially clear and agent decisions are made asynchronously. This, when coupled with the fact that agent decisions depend on the LOB and the LOB depends on agent decisions (there is a feedback loop) means that we require a mechanism to maintain market liquidity whilst at the same time maintaining a realistic spread, imbalance, etc. In other words, we need a market maker.

\citet{kumar2013implementing} implement an agent-based artificial exchange using the extended \citet{GlostenMilgrom1985} model of \citet{das2005learning}. This implementation cannot be considered truly reactive and bundles order matching with the model management in a sequential system thereby exhibiting high coupling, low concurrency but high cohesion. The \citet{GlostenMilgrom1985} model is a widely used and well understood market making model. However, we require a distribution for the true value of the stock while having a proportion of informed traders and a proportion of uninformed traders. The idea is that the market maker wants to know how to place the bids and asks based on the order flow from the traders. The model can then only work for toy cases where one can maintain the probability density estimate over the true value of the stock, thus allowing expectation to be taken. This is not possible in our simulation setting. 

\citet{das2005learning} extends the Glosten--Milgrom model by using an online algorithm that does not require tracking the true value of the stock. Rather, it applies Bayesian non-parametric methods to place Limit Orders (LOs) appropriately into the LOB. There is still a problem with the Glosten--Milgrom model, because it is a model with only one market maker that only provides liquidity at one level, and there is no method in determining the volume it should place. This needs to be written in by hand.

\citet{chakraborty2011market} provide an insightful derivation on the profitability of market making. This is central to the decision making made by any market maker. The setup is simple: at each time point the market maker places bids and asks around the mid-price up to a depth of $C_t$ at each time point $t$. The work provides an intuition on how market making can be profitable and thus self-sustaining. They show that when the local price movements is larger than the volatility from the change in opening and closing prices then market making can be profitable, {\it i.e.,} market making is profitable when there is a large amount of local price movement, but only a small net change in the price. They are quite clear in articulating the important difference between market making and statistical arbitrage. Market making places no directional bets and tries to minimises directional risk. This is a fundamental point that should not be lost when building financial market agent-based models.

\citet{AvellanedaStoikov2008} provide a stochastic control setup for market making. They make strong assumptions about the mid-price dynamics and utility function. The problem is their model only places orders with a volume size of one and it will only place limit prices on one level. This later example provides a natural conceptual split between the building of models for a single market making agent when designing trading algorithms to interact with a given market and its existing participants, as opposed to providing a minimalist specification of a market maker like interaction that can provide reasonably stable and realistic collective market dynamics. Here we are concerned with the later.

\section{Agent specifications}\label{sec:agentsetup}

There are several problems to overcome when shifting from sequential market clearing to a truly continuous double auction implementation. First, the agent rules can no longer depend on clearing prices from homogeneous time steps. The rules must be changed to an asynchronous framework where the agents decisions should depend on variables at any given time $t$. Second, there needs to be sufficient liquidity present, or at least interaction mechanisms in place to provide orders on both sides of the limit order book at any given time for the market to function. This is not required when applying sequential market clearing, but it is a necessary requirement in a continuous double auction.

Market makers usually fulfil the role of providing liquidity. The hallmark of market making is that they do not deliberately place directional bets but rather aim to profit off the spread while minimising directional risk \cite{chakraborty2011market}. The problem with market making models such as the Glosten--Milgrom model \cite{GlostenMilgrom1985} or more recent formulations into stochastic control problems \cite{AvellanedaStoikov2008,cartea2015algorithmic} is that the focus is only on the optimal placement of limit orders. These formulations do not guide the choice of the size of the order, nor do they cater for the placement of orders at multiple levels. Additionally, the formulation in a stochastic control problem setting does not fit easily within an agent-based model framework. Therefore, we avoid using market maker algorithms and follow \citet{leal2016rock} who used multiple high-frequency agents to mimic the behaviour of market makers in collective.

Concretely, our agent design largely follows \citet{leal2016rock} with several modifications to account for our continuous double auction setup. We also incorporate ideas from \citet{preis2006multi} and \citet{Alexandru2015} to create more realistic agent behaviours. The market consists of $N_{LT}$ Liquidity Takers (LT) that only submit market orders and $N_{LP}$ Liquidity Providers (LP) that only submit limit orders. The ratio of liquidity providers to liquidity takers is $N$:
\begin{equation} \label{eq:popratio}
N = \frac{N_{_{LP}}}{N_{_{LT}}}.
\end{equation}
The volume of each agent's order will be sampled according to a power law distribution with a density function given as:
\begin{equation}\label{eq:1}
    f(x)=
    \begin{cases}
    \frac{\alpha x_{\mathrm{m}}^{\alpha}}{x^{\alpha+1}} & \text{if }  x \geq x_{\mathrm{m}} \\
    0 & \text{if }  x<x_{\mathrm{m}}.
    \end{cases}
\end{equation}
Here the parameters $x_{\mathrm{m}}$ and $\alpha$ will depend on the state of the LOB and activation rules. \citet{Alexandru2015} samples the volume size according to an independent log-normal. However, from an empirical perspective, the unconditional distribution of the order size follows a power-law behaviour \cite{chakraborti2011econophysics1}. For the power law distribution, $x_{\mathrm{m}}$ is the minimum size of the volume order and a small $\alpha$ generates larger samples while a larger $\alpha$ generates smaller samples. These properties will become important when we couple the parameters to the state of the LOB and activation rules to achieve our desired agent behaviour.

\subsection{Liquidity takers \label{ssec:liquidity takers}}

We have two types of LT agents: ``fundamentalists" and ``chartists" (or trend-following agents) with a population of $i = 1,\ldots, N^f_{LT}$ and $i = 1,\ldots, N^c_{LT}$ respectively where the total population of liquidity takers is: $N_{LT} = N^f_{LT} + N^c_{LT}$. Each agent makes a decision to either buy or sell according to a random inter-arrival time governed by a truncated exponential distribution with mean intensity $\lambda^T$ where the intensity is bounded between $\lambda^T_{\min}$ and $\lambda^T_{\max}$ with time measured in seconds.

Each fundamentalist agents decision to buy (or sell) is given as
\begin{equation}\label{eq:2}
    D^f_{it} = 
    \begin{cases}
    \text{sell} & \text{if } f_{i} < m_t   \\
    \text{buy} & \text{if }  f_{i} > m_t.
    \end{cases}
\end{equation}
Here, $m_t$ is the mid-price at time $t$, $f_{i}$ is the fundamental value for the $i$th agent. The fundamental value for each of the fundamentalist is unique and fixed for the given trading day, specifically for that day's trading session. It is computed as 
\begin{equation}\label{eq:sigma}
    f_i = m_0 e^{x_i} \quad \text{where} \quad x_i \sim \mathcal{N}\left(0, \sigma^2\right).
\end{equation}
This is different from \citet{leal2016rock} where all agents have the same fundamental value at time $t$ and the fundamental value fluctuates between trading sessions.

Each chartist agents decision to buy (or sell) is given by:
\begin{equation}\label{eq:3}
    D^c_{it} = 
    \begin{cases}
    \text{sell} & \text{if } m_t < \Bar{m}_{it}   \\
    \text{buy} & \text{if }  m_t > \Bar{m}_{it},
    \end{cases}
\end{equation}
where $\Bar{m}_{it}$ is the exponential moving average (EMA) of the mid-price for the $i$th agent at time $t$. The EMA for each chartist is unique to themselves and is computed each time they make a decision. The moving average is updated according to $\Bar{m}_{it} = \Bar{m}_{it'} + \lambda \left( m_t - \Bar{m}_{it'} \right)$ where $t'$ is the time point where the $i$th agent made their last decision and $\lambda = 1 - e^{-\Delta t / \tau}$. Here $\Delta t = t - t'$ is the inter-arrival of the $i$th agent's current decision and previous decision and $\tau$ is the time constant for the $i$th agent computed as the mean inter-arrival of the agent's decision time. This allows our agents to base their decisions on variables at time $t$.

To create the market orders for LF agents to submit, we need to know both the direction of the order and the quantity of the order. The direction of the order is determined by \cref{eq:2,eq:3}, and the quantity of the order is sampled according to \cref{eq:1} with parameters determined according the magnitude of the activation rule and the state of the order book.

The lower bound $x_{\mathrm{m}}$ of the volume size is determined by the deviation between the agent's valuation or trading strategy relative to the current mid-price. We set
\begin{equation}\label{eq:volsize1}
x_{\mathrm{m}} =
\begin{cases}
    20 & \text{if } \left| f_{it} - m_t \right| \leq \delta m_t     \\
    50 & \text{if } \left| f_{it} - m_t \right| > \delta m_t,
\end{cases}
\end{equation}
for fundamentalists, and 
\begin{equation}\label{eq:volsize2}
x_{\mathrm{m}} =
\begin{cases}
    20 & \text{if } \left| m_t - \Bar{m}_{it} \right| \leq \delta m_t     \\
    50 & \text{if } \left| m_t - \Bar{m}_{it} \right| > \delta m_t,
\end{cases}
\end{equation}
for chartists.
Here the cutoff is a $\delta$ percentage of the current mid-price. This setup increases the aggression of the agent depending on how favourable the current price is relative to their valuation or trading strategy.

The second factor that determines the size of an agent's volume is the order book imbalance:
$$
\rho_t = \frac{v^b_t - v^a_t}{v^b_t + v^a_t},
$$
where $v^b_t$ and $v^a_t$ are the total bid and ask side volume of the LOB at time $t$. The shape parameter $\alpha$ is then:
\begin{equation}\label{eq:ascale1}
\alpha =
\begin{cases}
    1 - \rho_t / \nu & \text{for sell MO}   \\
    1 + \rho_t / \nu & \text{for buy MO},
\end{cases}
\end{equation}
where $\nu > 1$ to ensure that $\alpha$ is never equal to zero. The setup achieves two effects. First, if the contra side of the order book is thicker then an agent will submit larger orders to take advantage of the available liquidity. Second, if the same side of the order book is thicker then they will submit a smaller order to avoid excessive price impact.

\subsection{Liquidity providers \label{ssec:liquidity providers}}

Each LP agent makes a decision to place either a bid or ask limit order according to an random inter-arrival time governed by a truncated exponential distribution with mean intensity $\lambda^P$ that is bounded between $\lambda^P_{\min}$ and $\lambda^P_{\max}$ seconds. At each decision point the agent places either a bid or ask depending on the current order book imbalance $\rho_t$ at time $t$. An agent has a probability $\theta$ of placing an ask, and probability of $1-\theta$ for the placing of a bid where $\theta = \frac{1}{2}(\rho_t + 1)$. This design ensures that liquidity providers will on average provide liquidity to the side with less liquidity and thus stabilise the order book.

Once an agent has made the decision on which side of the order book to place the limit order, the placement of the limit order is given as:
$$
p_t =
\begin{cases}
    p^b_t + 1 + \lfloor \eta \rfloor & \text{for asks}   \\
    p^a_t - 1 - \lfloor \eta \rfloor & \text{for bids},
\end{cases}
$$
where $p^b_t$ and $p^a_t$ is the best bid and best ask at time $t$ respectively. Here $\eta$ is a random sample from a gamma distribution with a density given as
$$
f(x)=\frac{\beta^{k}}{\Gamma(k)} x^{k-1} e^{-\beta x},
$$
where $k$ is the shape parameter and $\beta$ is the rate parameter. The shape parameter $k$ is set to be the spread $s_t$ at time $t$,\footnote{The spread is the difference between the best ask and the best bid, {\it i.e.,} $s_t = p^a_t - p^b_t$.} whereas the rate parameter is set as
$$
\beta = 
\begin{cases}
    e^{-\rho_t/\kappa} & \text{for asks}   \\
    e^{\rho_t/\kappa} & \text{for bids},
\end{cases}
$$
at time $t$. This means the size of $\eta$ at time $t$ will on average be
\begin{equation}\label{eq:4}
    \bar{\eta} = k/\beta = 
    \begin{cases}
    s_t{e^{\rho_t/\kappa}} & \text{for asks}   \\
    s_t{e^{-\rho_t/\kappa}} & \text{for bids}.
    \end{cases}
\end{equation}
The above equation achieves several effects. 

First, when the order book is relatively balanced, the order placement will on average be near the best bid or ask. Second, when the order book is thicker on the contra side of the order book then the limit price placement will be more passive and will on average be placed further away from the best bid price to capture the strategic effect \cite{Alexandru2015}. Lastly, when the order book is thicker on the same side of the order book then the limit price placement will be more aggressive and will on average be placed between the spread to capture the competition effect \cite{Alexandru2015}. Here $\kappa > 0$ is a parameter that intensifies or relaxes the strategic or competition effect. When $\kappa$ is small the effects are intensified and when $\kappa$ is large the effects are relaxed. All limit orders have a time in force of $\gamma$ seconds.

\begin{remark}
Initially we sampled $\eta$ from an exponential distribution following \citet{preis2006multi} but with a mean of $\bar{\eta}$ from \cref{eq:4}. However, we found that the exponential distribution resulted in the spread closing too quickly even though the sample will on average be placed at the best bid/ask (provided $\rho_t \approx 0$). This is because the exponential distribution has more mass below the mean so that the limit price placement is more likely to be between the spread. This will reduce the spread for the next limit price placement and results in a feedback loop that closes the spread. The gamma distribution slows down the collapse which gives liquidity takers a chance to push back the spread.
\end{remark}

The volume of each limit order is sampled according to \cref{eq:1} with the lower bound of the volume size as $x_n = 10$ and the shape parameter as
\begin{equation}\label{eq:ascale2}
\alpha =
\begin{cases}
    1 - \rho_t / \nu & \text{for asks}   \\
    1 + \rho_t / \nu & \text{for bids}.
\end{cases}
\end{equation}
This setup achieves two effects. First, when the order book is thicker on the contra side of the order then the volume of the limit order will be large. Second, when the order book is thicker on the same side as the order then the volume of the limit order will be small. The motivation behind this choice was to ensure that imbalances are swiftly addressed to ensure that we have a stable order book.

\section{System implementation} \label{sec:implementation}

\begin{figure}[htb]
    \centering
    \begin{tikzpicture}[node distance=2cm, scale = 0.6]
        \node (cointossx) [rectangle,rounded corners,minimum width=3cm,minimum height=1cm,text centered,draw=red,thick] {CoinTossX};
        \node[align=center, xshift=2cm] (listener) [trapezium,trapezium left angle=70,trapezium right angle=110,minimum width=3cm,minimum height=1cm,text centered,draw=purple,thick,right of=cointossx] {Market Data\\Listener};
        \node (lob) [diamond,minimum width=3cm,minimum height=1cm,text centered,draw=black,thick,below of=listener,align=center,yshift=-1cm] {Update\\LOB\\States};
        \node (simulation) [diamond,minimum width=3cm,minimum height=1cm,text centered,draw=green,thick,below of=cointossx,align=center,yshift=-1cm] {Agent\\Decision\\Rules};
        \draw [thick,->,>=stealth,dotted] (cointossx) -- node[anchor=south] {feed} (listener);
        \draw [thick,->,>=stealth] (listener) -- (lob);
        \draw [thick,->,>=stealth] (lob) -- (simulation);
        \draw [thick,->,>=stealth,dotted] (simulation) -- node[anchor=east] {orders} (cointossx);
        \draw[densely dotted,thick] (3,2) -- (11,2) -- (11,-7.5) -- (-3,-7.5) -- (-3,-2) -- (3,-2) -- (3, 2);
        \node at (10,-6) {MMS};
    \end{tikzpicture}
    \caption{Schematic representation of the separation of the Model Management System (MMS) which replaces the functionality of the Execution Management Systems (EMS) and the Order Management System (OMS) in a conventional trading environment. The Matching Engine (ME) is implemented using CoinTossX \cite{sing2017cointossx}. The ME is loosely coupled to the MMS. The MMS accesses messages from the data feed by listening to a port using a socket. Orders are injected as message strings and pushed to a port. \label{fig:flowchart}}
\end{figure}
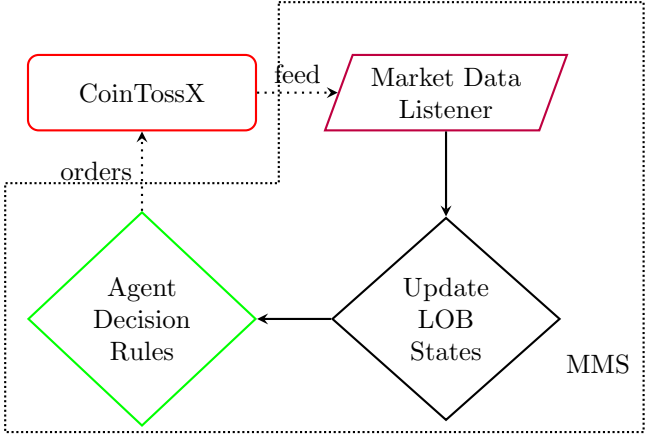

\Cref{fig:flowchart} outlines how the various components of the model fit together. The Matching Engine (ME) is its own independent piece of software which is coupled to the Model Management System (MMS) by the data feed and submission of orders. The model management system consists of the market data listener and the agent decision rules. Here the market data listener builds its own LOB by listening to the data feed from the matching engine and updating the LOB whenever a message comes through. The updated LOB can be accessed by all the agents which then make a decision and submit their order to the matching engine.

\subsection{Matching engine}

To achieve low-coupling and high-cohesion between the Matching Engine (ME) and our agent-based model we required further development of the matching engine code and the order management and model management framework.\footnote{Here cohesion refers to the degree to which a part of a code base forms a logically single, atomic unit. Here coupling refers to the degree to which a single unit is independent from others, {\it i.e.,} it is related to the number of connections between two or more units.} In our previous work we had a high-coupling between the matching engine and our simulation. This was because the state of the LOB was observed by requesting snapshots using a back-end functionality from the matching engine \cite{arxiv2021hawkes}. The back-end functionality was designed to help debug the matching engine when it was being built and tested; it was not intended to be used as part of the simulation workflow functionality.  

The system has now been separated with order submissions/confirmations and executions being forwarded in real time over a network via UDP from the matching engine, {\it i.e.,} we have a continuous data feed.\footnote{User Datagram Protocol (UDP) is a lightweight, connectionless (unreliable) but fast protocol where data is transmitted irrespective of whether there is a listener to receive the message or not.} These messages are then continuously captured and managed by a dedicated order management shim on the modelling framework side.

For the continuous data feed we encode and decode messages in a binary format such that transmission of messages over the network remains fast. This is currently a prototype for more advancements to be made in the flexibility of CoinTossX's API with regards to other programming languages. 
\subsection{Market-data listener}

The market data listener is focused on listening for messages on the data feed from CoinTossX and processing these messages to build a separate LOB for the agents that is independent of the matching engine. This mirrors the workflow of real market participants who collate and manage their own representations of the limit order book for decision making. Sample messages received from CoinTossX are illustrated in \ref{app:data}.

The LOB is created from the messages using separate dictionaries for the ask side and bid side. Each order is stored as key-value pairs where the keys are given by the order IDs assigned from CoinTossX while the value for the key contains the limit price and volume. With this setup, updating the LOB is simply a matter of adding new orders and removing or cancelling orders by cross-referencing order IDs given in transactions or cancellations (see pseudocode \ref{algo:lob state} for maintaining the LOB). Note that the LOB built in the MMS can only keep track of the price priority and not the time priority. The full price-time priority is maintained in the matching engine. Therefore, the order ID is the most important component to help maintain the LOB in the MMS. This reflects what happens in real financial markets.

Once the LOB has been updated we extract some important quantities upon which our agent decisions depend on, such as: the best bid $p^b_t$ and ask $p^a_t$, the mid-price $m_t$, the spread $s_t$ and the order book imbalance $\rho_t$. It should be noted that when the bid side of the LOB is empty, the best bid price $p^b_t$ retains the price of the previous best bid before the bid side was emptied as a reference price. When the ask side of the LOB is empty, $p^a_t$ retains the price of the previous best ask before the ask side emptied as its reference price.

The spread $s_t$ is computed as $p^a_t - p^b_t$ and the mid-price $m_t$ as $\frac{p^a_t + p^b_t}{2}$, even if the best bid or ask are merely indicative (from the last existing reference price). The imbalance can always be computed even if one side is empty as it will simply take on $1$ for an empty ask side or $-1$ for an empty bid side. The same logic is applied when both sides of the LOB are empty, except that $\rho_t$ will be set to $0$ in this particular case. This ``fail-safe" setup for an empty LOB scenario makes it easy to initialise the simulation because the simulation can begin with just an indicative best bid and ask. This can avoid unnecessary volatility auctions which are currently not accounted for by our agent specifications.

\subsection{Agent decision rules}

The order sent by the agent is determined by the decision rules discussed in \Cref{sec:agentsetup}. Since agent actions depend on state variables of the LOB at time $t$, all the agents have access to the same LOB constructed from the market data listener but maintained in the model management system that mimics the role of order management system in the traditional trader workflow.

The arrival of decision times for all agents follow a truncated exponential. Therefore, we can simulate and sort all the decision time points in chronological order before starting the simulation. This means we can login to CoinTossX with a single client to manage and submit the orders for all the agents. At each decision point we identify the type and the specific agent and create the order for the agent according to the appropriate rules using the current state of the LOB. This is again a model design convenience specific to our particular model as managed from the MMS.

The simulation does not have to occur in real time and can be significantly sped up because we know all the decision times beforehand and we do not have any reactive agents.\footnote{By reactive agents we mean agents who act on events, {\it i.e.,} agents listen for events and then act on them to define an event time.} The appropriate event times can be appended to the trade and quote data after the simulation. The only requirement to achieve this is the appropriate ordering of events so that the price-time priority of orders can be maintained within the matching engine.

\subsection{Compute times and hardware} \label{ssec:computetime}

The Microsoft Azure \cite{MSAZure} hardware configuration used and indicative times for each of the key steps of the simulation framework are given in \Cref{tab:hardware}. All times mentioned in \Cref{tab:hardware} are dependent on the number of orders submitted and therefore the number of trader agents.
\begin{table}[h]
    \centering
    \begin{tabular}{p{2.75cm}p{5.25cm}} 
    \toprule
    Hardware & Specification \\ 
    \midrule
    Virtual Machine & Standard A4m V2 Azure VM \cite{MSAzureVM} \\
    Operating System & 64-bit Linux Ubuntu 18.04-LTS\\
    RAM & 32GB \\
    CPU & $4 \times$ Intel Xeon CPU E5-2673 v3\\
    & @ 2.4GHz (32 threads) \\ 
    \midrule
    Computation & Run-time \\ 
    \midrule
    1-hour of Trading & 41.179s \\
    Sensitivity Ana. & 35.746hrs \\
    Calibration & 10.880hrs \\ 
    \bottomrule
    \end{tabular}
    \caption{Hardware configuration and indicative computation times. The average hour of trading for a single stock results in between a total of 4800 and 6600 orders. This includes, on average, between 2200 and 3100 limit orders, between 2000 and 2800 cancellations and between 350 and 450 trades.} \label{tab:hardware}
\end{table}

\subsection{Miscellaneous considerations}

In the Johannesburg Stock Exchange (JSE) a volatility auction call session can be triggered during a continuous double auction when an instrument's circuit breaker tolerance has been breached \cite{JSEvolauction}. This means that if the execution price is larger than a 10\% deviation from the dynamic reference price then a volatility auction will be triggered. The execution price is the price of the best bid or ask that is crossed when a market order arrives while the dynamic reference price is the deepest price level executed from the last transaction \cite{sing2017cointossx}.

Our agent specification does not account for the appropriate agent behaviour during an auction. Therefore, we impose a restriction on all liquidity takers where they will not submit their market order if the order will lead to an execution price with a deviation larger than 10\% of the current dynamic reference price in order to avoid triggering a volatility auction.

Lastly, liquidity takers will not submit their market order if the contra side of their order is empty. We do this to simplify the data cleaning process. This is because when a market order is submitted to an empty order book the data feed returns a string but with no orders executed. This means that sending or not sending the order will lead to the same result.

\subsection{Simulation}

To initialise all simulations we fix the number of liquidity takers with $N^f_{LT} = 1$ and $N^c_{LT} = 1$ and only vary the number of liquidity takers during calibration and sensitivity analysis. The LOB state of the best ask, best bid, spread, imbalance $(a_t, b_t, s_t, \rho_t)$ is arbitrarily set as (9950, 10050, 100, 0) and we start with an empty LOB.

\begin{figure*}[!htb]
    \centering
    \includegraphics[width=\textwidth]{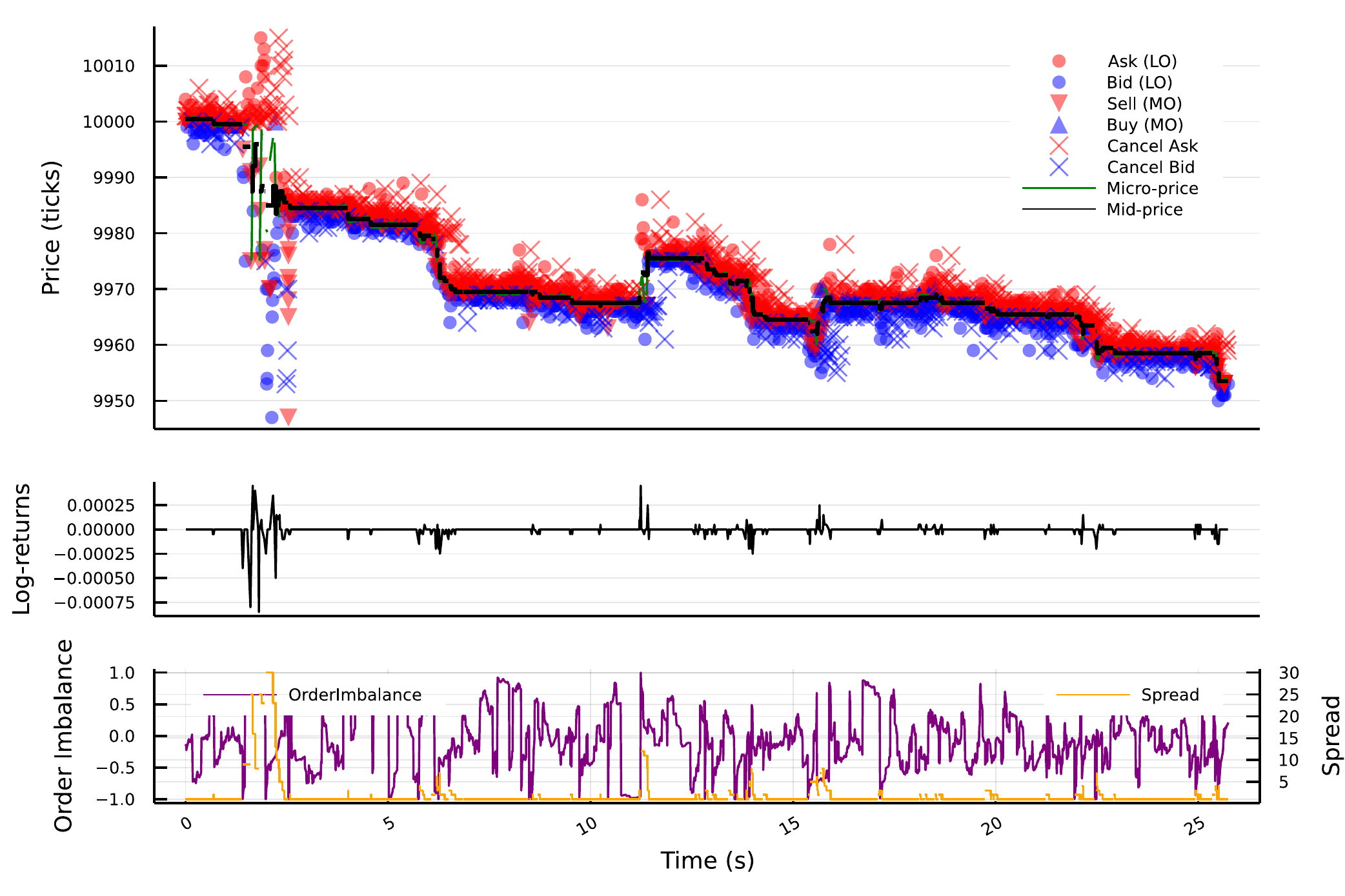}
    \caption{Half a minute of simulated price time-series from calibrated parameters. Orders placed at all levels of the LOB are visualised with markers for limit, market and cancellations of buy and sell orders. The progression of returns, spread and LOB volume imbalance are also visualised over the same time period. \label{fig:simulation}}
\end{figure*}

Simulations carried out on calibrated parameters had the most consistent behaviour compared to ideal parameters found through sensitivity analysis. That is, parameters selected based on qualitative evaluations of moments of simulated price time-series, produced simulations with little to no liquidity taker activation's. This resulted in almost no price movement for longer simulations.\footnote{Due to the long simulation times with the Julia-Java wrapper, memory usage and speed became important. This meant JVM arguments for garbage collection and memory allocation were required.} Figure \ref{fig:simulation} provides the simulated price time-series obtained from calibrated parameters for a 30 second interval.

\section{Sensitivity analysis \label{sec:sensitivity}}

We conduct a sensitivity analysis on the moments of both simulated mid-price and micro-price. This serves to demonstrate the effect of different price time-series on the estimates of simulated moments. Our choice of moments and statistics corresponds to those recommended by \citet{gilli2003global} which are robust enough to reflect the properties of financial data and flexible enough to discriminate between parameter configurations \cite{winker2007objective}. Therefore, we adopt the following moments and statistics for being the most representative measures of the stylised facts of financial data: mean, standard deviation, kurtosis, Kolmogorov-Smirnov (KS) statistics, Geweke-Porter-Hudak estimator (GPD), augmented Dicky-Fuller (ADF), a combination of GARCH(1,1) parameters, and a Hill estimator.\footnote{The modified Hill estimator is computed by numerically solving $\frac{1}{l - r + 1} \sum_{j = r}^{l} \ln{(X_j)} = \frac{1}{\alpha} + \frac{\ln{(X_l)} X_l^{-\alpha} - \ln{(X_r)} X_r^{-\alpha}}{X_l^{-\alpha} - X_r^{-\alpha}}$ where $X_l$ and $X_r$ are the minimum and maximum return values respectively \cite{nuyts2010inference}.} These are given in Table \ref{tab:calibrationmoments}. 

\begin{table*}[h]
\centering
\begin{tabular}{p{4.5cm} p{12.5cm}}
\toprule
Targeted Empirical Feature & Estimators and Motivation \\
\toprule
Distribution shape & The mean $\mu$, standard deviation $\sigma$, kurtosis $\mu_4$, and Kolmogorov-Smirnov (KS) statistic of log-returns are chosen to represent the overall shape of the data distribution. The KS statistic compares the empirical distribution of the actual returns, to that of the simulated or bootstrapped returns obtained from the model. The actual returns obtain a KS statistic of 0 as they are identically distributed to themselves. In the context of both the calibration and the sensitivity analysis, the KS statistic was computed relative to the empirical distribution.\\
\midrule
Long-range dependence & The Geweke and Porter-Hudak (GPH) estimator: \citet{geweke1983estimation}, and \citet{lillo2004long} provide a measure of the long-range dependence of the absolute log returns by estimating the memory parameter $d \in [-0.5, 0.5]$ of the log-periodogram.\\
\midrule Serial auto-correlation & Augmented Dickey-Fuller (ADF) statistic: ADF measures of the extent of the random walk property of log returns by testing the null hypothesis that a unit root is present in the time series. A large negative test statistic is more evidence against the null hypothesis. \\
\midrule
Long-memory & The empirical Hurst exponent (H): The Hurst exponent relates to the auto-correlations of the time series, and its rate of decay for increasing lags. A value in the range 0.5 and 1 indicates a time series with long-term positive auto-correlation (momentum) while a value in the range 0 and 0.5 is indicative of long-term negative auto-correlation (mean reversion). This is indicative of long-memory effects. \\
\midrule
Short-range dependence & The sum of the two GARCH(1,1) parameters is used as a measure of short-range dependence. The reason for taking the sum is that this value is much more robust than using either one of the two parameters in the GARCH model alone.\\
\midrule
Power-law tail & An ``improved'' Hill estimator (HE): \citet{nuyts2010inference} argue for the estimation of the tail index of a power-law distribution as an additional measure of the power behaviour and tailedness of the distribution over and above kurtosis. The use of this modified estimator is in response to the known issues and inefficiencies of the standard estimator.\\
\bottomrule
\end{tabular}
\caption{Estimators and moment parameters used for the calibration and sensitivity analysis along with the properties they are targeting. See sections \ref{sec:sensitivity} and \ref{sec:calibration}. The moments are used in the Nelder-Mead (NM) method with Simulated Minimum Distance (SMD) for the indicative calibration because of the methods simplicity, speed and reliability.} \label{tab:calibrationmoments}
\end{table*}

Overall, this results in a combination of 9 moments and statistics being used to characterise the statistical properties of the simulations.

To visualise the impact of pairs of parameters on the statistical properties of the simulated mid and micro price, we consider combinations of a range of reasonable parameter values. That is, we iteratively fix each parameter and vary the others in their ranges such that we obtain moment surfaces for all combinations of parameter values. We considered a range of five different values for each parameter, resulting in 3125 different combinations. Each unique parameter value thus has 625 replications over which to aggregate moments.

Given the large number of possible parameter configurations and that the model management system is not coupled to the matching engine system, some care must be taken to ensure that separate back-to-back order submission periods are efficient, independent and do not result in unexpected behaviour (both at a software level and simulation output level). For example, we require that each simulation begins with an empty limit order book such that orders from previous simulations don't carry over. One way to achieve this would be to initiate new continuous trading sessions at set time intervals. Simulation time, however, is not fixed and this could cause orders to be submitted in the wrong trading session. For this reason we manually clear the off-heap storage and LOB each time the client logs out.

Figures \ref{fig:sensitivity boxplot midprice}, \ref{fig:sensitivity boxplot microprice} and \ref{fig:moment surfaces} in \ref{app:sensitivity analysis results} provide high-level summaries of the individual effects of parameter values on the moments of simulated log-return time-series.

Kurtosis was found to be the least reliable moment since extreme price changes during the initialisation period would have resulted in a large kurtosis value even if the price remained constant for the rest of the simulation.\footnote{During calibration the instability of certain moments is accounted for by the inverse covariance matrix of empirical moments by down-weighting the effect of those moments on the objective function (refer to \Cref{sec:calibration}).} Moments that were robust and good indicators of realistic price series include the GPH statistic, the Hurst exponent, and the Hill estimator. Due to the nature of the model, the Hurst exponent remained relatively constant and above 0.5 across parameters and simulations indicating a tendency towards momentum behaviour. Assigning a value of close to zero for the $\sigma$ parameter of fundamentalists resulted in mostly mean-reverting behaviour with a Hurst exponent close to 0.5.

\begin{figure*}[ht]
    \centering
    \includegraphics[width=.75\textwidth]{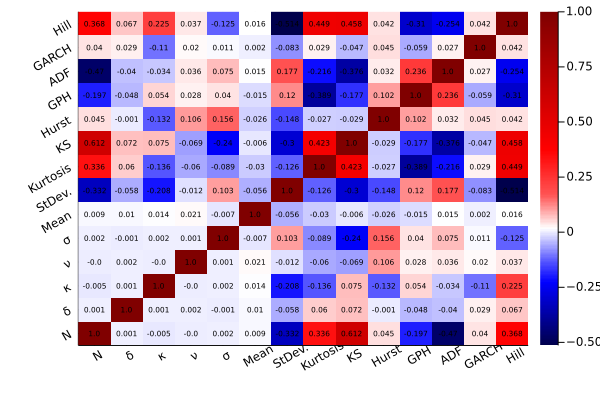}
    \caption{Bootstrapped correlations between individual parameter-moment combinations calculated on simulated micro-price time-series obtained from the sensitivity analysis. The parameter descriptions are provided in Table \ref{tab:freeparameters} and the moment descriptions in Table \ref{tab:calibrationmoments}. \label{fig:momentparameter correlation}}
\end{figure*}

On average, most parameters by themselves do result in significant differences in simulated moments across their range of values. On the other hand, individual differences in parameters $\kappa$ and $N$ were found to result in slightly different simulated moments across their range of values on average. The strong dependence of simulated moments on $N$ is expected as this parameter controls the number of liquidity providers. Smaller values of $\kappa$ exhibited more leptokurtic log-returns which captures the intended effect in the model definition --- small values of $\kappa$ intensifies competition while larger values relax competition. Another observation worth noting is the relatively stable Hurst exponent which, with the exception of $N$, remains consistently above 0.5 --- indicating a time-series exhibiting largely momentum behaviour.

Considering different parameter combinations together produced simulated moments that were significantly different from those obtained by averaging the marginal effects of individual parameters over all other parameter combinations. This can been by the kurtosis, Hurst exponent and Hill estimator moment surfaces in \Cref{fig:moment surfaces} for combinations of $\delta$ \& $\kappa$ and $N$ \& $\nu$. Certain moment surfaces in \Cref{fig:moment surfaces} differ depending on the use of mid-price or micro-price log-returns.

From Figure \ref{fig:momentparameter correlation}, as expected, the kurtosis, Kolmogorov-Smirnov statistic and Hill estimator are strongly correlated to the ratio of liquidity providers to liquidity takers while the ADF statistic and standard deviation of log-returns are negatively correlated. Similarly, with $\kappa$ controlling the order-placement depth, it is positively correlated with the Hill estimator and negatively correlated with the standard deviation of log-returns.

\section{Calibration \label{sec:calibration}}

The three most common calibration strategies are (i) the maximum likelihood approach which is applicable to models with closed form solutions, (ii) the sequential Monte-Carlo approximate Bayesian computation method,\footnote{SMC-ABC was applied to the \citet{preis2006multi} model in \cite{goosengebbie2020calibrating} with the surprising result that frequentist methods were sufficient.} and (iii) the method-of-moments (MM) which often employs variations of simulated minimum distance (SMD) methods. The MM approach has been criticised for requiring the selection of an arbitrary set of moments and for creating a rough objective surface with many local minima and for only generating point estimates as opposed to credibility intervals and posterior distributions.\footnote{For an extensive comparison of the many different calibration techniques refer to \citet{platt2020comparison}.} Nonetheless, ignoring parameter inference, we apply the MM SMD method for its simplicity, speed and reliability when there are more moments than parameters and we have significant computational time and resource constraints. As such our calibration should be considered indicative rather the robust. From the full set of model parameters a small subset of 5 free parameters was chosen: $\ve \theta = \{ N, \delta, \kappa, \nu, \sigma \}$ (see Section \ref{sec:agentsetup} and Figure \ref{fig:abm state flow}). This set of 5 parameters are described in Table \ref{tab:freeparameters} and are the parameters to be calibrated.

The list of fixed parameters that control other simulation characteristics are as follows. The inter-arrival rate parameter with minimum and maximum sampling intervals for liquidity takers ($\lambda^{T}$, $\lambda^{T}_{\min}$ and $\lambda^{T}_{\max}$ respectively) and liquidity providers ($\lambda^{P}$, $\lambda^{P}_{\min}$ and $\lambda^{P}_{\max}$ respectively) control for any low-frequency or high-frequency trading behaviour and affect, among other things, the amount of liquidity in the order book. These parameters are, however, redundant in the presence of parameters $\nu$ and $\delta$ which also control liquidity. Here simulations were not conducted in real time, but for computational convenience, event times were appended to the data after submission. Lastly, there was the power-law order-size cut-off parameter $x_m$ as well as the time-to-arrival of cancellations that were held fixed which also affected the build up of liquidity in the order book. These components were kept fixed to alleviate calibration problems that may occur with redundant behaviour specifications as a result of conflicting and dependent parameters.

\begin{table}[htb]
    \centering
    \begin{tabular}{lp{6.5cm}}
    \toprule
    Param. & Description \\
    \midrule
    $N$ & The population ratio of liquidity takers to liquidity providers (see \cref{eq:popratio}).    \\
    $\delta$  & Volume size aggression multiplier as a percentage of mid-price (see \cref{eq:volsize1,eq:volsize2}).   \\
    $\kappa$ & Placement depth multiplier (see \cref{eq:4}). \\
    $\nu$ & Scaling factor for power-law volume order size (see \cref{eq:ascale1,eq:ascale2}). \\
    $\sigma$ & Fundamentalists' value perception uncertainty for the trading day (see \cref{eq:sigma}). \\
    \bottomrule
    \end{tabular}
    \caption{Free model parameters (for calibrated values see Table \ref{tab:parameter inference}).} \label{tab:freeparameters}
\end{table}

The task of calibrating intraday ABMs in particular has been proven to be challenging for a number of reasons \cite{plattgebbie2016problem}. Firstly, model parameters often demonstrate degeneracies and have a haphazard effect on the objective function which leads to consecutive calibration experiments with significantly different parameters. Certain parameters tend to dominate all the others and it is often difficult to identify the role and dependencies between the various parameters. For the models we have considered the degeneracies are hypothesised to find much of their origin in the realistic matching processes, and in the particular model and its price setting mechanism \cite{plattgebbie2018can}. In our case, agent decision rules are explicitly encoded, resulting in fewer parameters to calibrate and possibly a more unique set of calibrated parameters. This also ensures that the number of free parameters is less than the number of moments used to construct the objective function.

For the problem of calibrating models with as few parameters as this one, we adopt the frequentist/method-of-moments simulated minimum distance (SMD) approach where the objective function is chosen to take into account the stylised facts of financial data \cite{winker2007objective}. The objective function to be minimised is thus the weighted sum-of-squares of estimation errors between simulated and empirical moments. The moments to be estimated are those mentioned in \Cref{sec:sensitivity} along with the mean, standard deviation and Komogorov-Smirnov statistic which are estimated on the micro-price log-returns.\footnote{Tick-by-tick mid-prices were found to unreliable when trying to identify distributional properties through moments.} Denoting $\bm{m^e} = [m_1^e, \hdots, m_k^e]^{\prime}$ and $[\bm{m^s} \mid \bm{\theta}] = [m_1^s, \hdots, m_k^s]^{\prime}$ as the vector of empirical moments of real and simulated data respectively, we minimise the quadratic function
$$
\min \limits_{\bm{\theta} \in \bm{\Theta}} f(\bm{\theta}) = G(\bm{\theta})^{\prime}\bm{W}G(\bm{\theta}),
$$
with
$$
G(\bm{\theta}) = \frac{1}{I} \sum_{i = 1}^{I}{(\bm{m_i^e} - [\bm{m_i^s} \mid \bm{\theta}])},
$$
where $\bm{\theta}$ is the vector of parameters, $\bm{\Theta}$ is the space of feasible parameters and $I = 5$ is the number of replications/simulations used in estimating the average moments and statistics. The matrix of weights $W$, given by the inverse covariance matrix of empirical moments ($\Cov^{-1}[\bm{m^e}]$), takes into account the joint distribution of moments and assigns larger weights to those moments with the least uncertainty and that are more robust \cite{winker2007objective}.\footnote{The weight matrix $\bm{W}$ is estimated by applying a moving block bootstrap to the time series with a window of size 2000. As opposed to sampling individual returns, sampling bootstrap blocks preserves the auto-correlations of the time-series. For each block/window we re-sample with replacement until we obtain 1000 samples, each of which is equal to the length of the original data set. The moments and statistics are then calculated on each of these samples such that we can obtain the covariance between them. The condition number of the weight matrix $\bm{W}$ is $8.375811152908104 \times 10^{19}$ computed as $||W^{-1}|| \cdot ||W||$.}

To alleviate the issue of a globally non-convex objective surface, we employ the Nelder-Mead with Threshold Accepting (NMTA) heuristic optimisation routine \cite{gao2012implementing,gilli2003global,nelder1965simplex} (refer to \ref{app:nmta}). The data on which the model is calibrated is chosen to be 1-hour of Naspers level-1 trade-and-quote data from the JSE.

From the sensitivity analysis we generate chains of parameters $\bm \theta$ from which we can compute the moments $\bm{m}^s$; each vector of moments is then associated with particular vectors of parameters. From the sensitivity analysis we can the compute the exposure $\bm{B}$ (here a $5 \times 9$ matrix) of parameters to the moments.\footnote{Here $B_{ij} = \Cov[\theta_i,m^s_j]/\Var[m^s_j]$ for $i$th parameter and $j$th moment in the exposure matrix $\bm{B}$.} Then from the bootstrapped moment covariance matrix $\Sigma_{m^e} = \Cov[\bm{m^{e}}] = \bm{W}^{-1}$ and the approximated parameter exposures to the moments $\bm{B}$, we can construct an indicative covariance for the parameters: $\Sigma_{\theta} = \bm{B}' \Sigma_{m^{e}} \bm{B}$. The diagonal of this gives us indicative sample variances: $\sigma^2_{\theta} = \diag(\Sigma_{\theta})$. Parameters obtained from calibration and sensitivity analysis are shown in Table \ref{tab:parameter inference}.

To obtain the indicative error bars on simulated moments in Table \ref{tab:moment inference}, we obtain moment variances from the diagonal of the bootstrapped moment covariance matrix $\Sigma_{m^e}$ and compute confidence intervals as $\bm{m}^s \pm 1.96 \times \sqrt{\diag(\Sigma_{m^e})}$.

\begin{table}[htb]
    \centering
    \begin{tabular}{lccc}
    \toprule
        Parameter & {$\hat{\theta}_{0.025\%}$}& {$\hat{\theta}$}& {$\hat{\theta}_{0.975\%}$} \\ \midrule
        $N$ & \DTLfetch{theta}{Variable}{Nh}{Lower} & \DTLfetch{theta}{Variable}{Nh}{Estimate} & \DTLfetch{theta}{Variable}{Nh}{Upper} \\
        $\delta$ & \DTLfetch{theta}{Variable}{Delta}{Lower} & \DTLfetch{theta}{Variable}{Delta}{Estimate} & \DTLfetch{theta}{Variable}{Delta}{Upper} \\
        $\kappa$ & \DTLfetch{theta}{Variable}{Kappa}{Lower} & \DTLfetch{theta}{Variable}{Kappa}{Estimate} & \DTLfetch{theta}{Variable}{Kappa}{Upper} \\
        $\nu$ & \DTLfetch{theta}{Variable}{Nu}{Lower} & \DTLfetch{theta}{Variable}{Nu}{Estimate} & \DTLfetch{theta}{Variable}{Nu}{Upper}  \\
        $\sigma$ & \DTLfetch{theta}{Variable}{Sigma}{Lower} & \DTLfetch{theta}{Variable}{Sigma}{Estimate} & \DTLfetch{theta}{Variable}{Sigma}{Upper} \\ \bottomrule
    \end{tabular}
    \caption{Calibrated parameters along with 95\% confidence intervals where parameter variances are obtained from sensitivity analysis parameter chains. See Table \ref{tab:freeparameters} for the parameter descriptions.} \label{tab:parameter inference}
\end{table}
\begin{table}[htb]
    \centering
    \begin{tabular}{lcccc}
    \toprule
        Moment & {${\hat{m}^s}_{0.025\%}$} & {$\hat{m}^s$} & {${\hat{m}^s}_{0.975\%}$} & {$\hat{m}^e$} \\ \midrule
        Mean & \DTLfetch{m}{Moment}{Mean}{Lower} & \DTLfetch{m}{Moment}{Mean}{Estimate} & \DTLfetch{m}{Moment}{Mean}{Upper} & \DTLfetch{m}{Moment}{Mean}{Empirical} \\
        StDev. & \DTLfetch{m}{Moment}{StandardDeviation}{Lower} & \DTLfetch{m}{Moment}{StandardDeviation}{Estimate} & \DTLfetch{m}{Moment}{StandardDeviation}{Upper} & \DTLfetch{m}{Moment}{StandardDeviation}{Empirical} \\
        Kurtosis & \DTLfetch{m}{Moment}{Kurtosis}{Lower} & \DTLfetch{m}{Moment}{Kurtosis}{Estimate} & \DTLfetch{m}{Moment}{Kurtosis}{Upper} & \DTLfetch{m}{Moment}{Kurtosis}{Empirical} \\
        KS & \DTLfetch{m}{Moment}{KolmogorovSmirnov}{Lower} & \DTLfetch{m}{Moment}{KolmogorovSmirnov}{Estimate} & \DTLfetch{m}{Moment}{KolmogorovSmirnov}{Upper} & \DTLfetch{m}{Moment}{KolmogorovSmirnov}{Empirical} \\
        Hurst & \DTLfetch{m}{Moment}{Hurst}{Lower} & \DTLfetch{m}{Moment}{Hurst}{Estimate} & \DTLfetch{m}{Moment}{Hurst}{Upper} & \DTLfetch{m}{Moment}{Hurst}{Empirical} \\
        GPH & \DTLfetch{m}{Moment}{GPH}{Lower} & \DTLfetch{m}{Moment}{GPH}{Estimate} & \DTLfetch{m}{Moment}{GPH}{Upper} & \DTLfetch{m}{Moment}{GPH}{Empirical} \\
        ADF & \DTLfetch{m}{Moment}{ADF}{Lower} & \DTLfetch{m}{Moment}{ADF}{Estimate} & \DTLfetch{m}{Moment}{ADF}{Upper} & \DTLfetch{m}{Moment}{ADF}{Empirical} \\
        GARCH & \DTLfetch{m}{Moment}{GARCH}{Lower} & \DTLfetch{m}{Moment}{GARCH}{Estimate} & \DTLfetch{m}{Moment}{GARCH}{Upper} & \DTLfetch{m}{Moment}{GARCH}{Empirical} \\
        Hill & \DTLfetch{m}{Moment}{Hill}{Lower} & \DTLfetch{m}{Moment}{Hill}{Estimate} & \DTLfetch{m}{Moment}{Hill}{Upper} & \DTLfetch{m}{Moment}{Hill}{Empirical} \\ \bottomrule
    \end{tabular}
    \caption{Comparison between empirical moments and moments estimated on tick-by-tick simulated micro-price returns along with 95\% confidence intervals. See Table \ref{tab:calibrationmoments} for the description of the calibration moments.} \label{tab:moment inference}
\end{table}

\section{Exploratory Data Analysis}

It is difficult to compare the ``goodness'' of one ABM to another. Such comparisons are usually based on the model's ability to replicate empirical observations which involves subjective comparisons given the qualitative nature of most stylised facts. There are a wide variety of financial ABM's that are able to replicate common empirically observed facts. However, \citet{plattgebbie2016problem} demonstrate possible inadequacies of a stylised fact-centric approach to model validation and stress the importance of rigorous calibration methods as well as qualitative measures relating to model parsimony and the ability of the model to make predictions or produce empirical phenomena not directly encoded in, or used in the estimation process. It is also usually the case that a unique set of parameters for the calibration of a specific model to a time series does not exist. Without a unique set of parameters allowing us to reproduce the properties of financial time series drawn from a particular market, we cannot argue that introducing and observing changes in the model truly reflects the same changes in the market.
A good model should both be able to replicate the stylised facts of financial markets and have parameters that behave in clear ways and do not have insignificant effects on the resulting behaviour of the simulation.

That said, given the uniqueness of different markets, qualitative empirical-versus-simulation comparisons remain the best validation methods in computational agent-based models. We choose an extensive list of stylised facts that are constraining enough to demonstrate the realism of our simulations (see \ref{app:stylisedfacts} for a shortened list of financial market stylised and empirical facts).

\subsection{Stylised facts}

The stylised facts presented hereafter are computed on micro-prices on a tick-by-tick basis and are compared between simulated and empirical data. The micro-price is updated whenever an event changes the best bid or ask. The returns are computed as price fluctuations:
$$
r_{t_k} = \ln\left( S_{t_{k+1}} \right) - \ln\left( S_{t_{k}} \right),
$$
where $S_{t_{k}}$ is the micro-price as time ${t_{k}}$. The stylised facts included are by no means exhaustive but rather include the most common of these: order-flow auto-correlation, log-return and extreme log-return distributions and log-return auto-correlation. In addition to these we compute the depth-profile curves on simulated LOB data.

\begin{figure*}[!htb]
    \centering
    \subfloat[CoinTossX - calibrated parameters\label{figa:return distribution}]{\includegraphics[width=.5\textwidth]{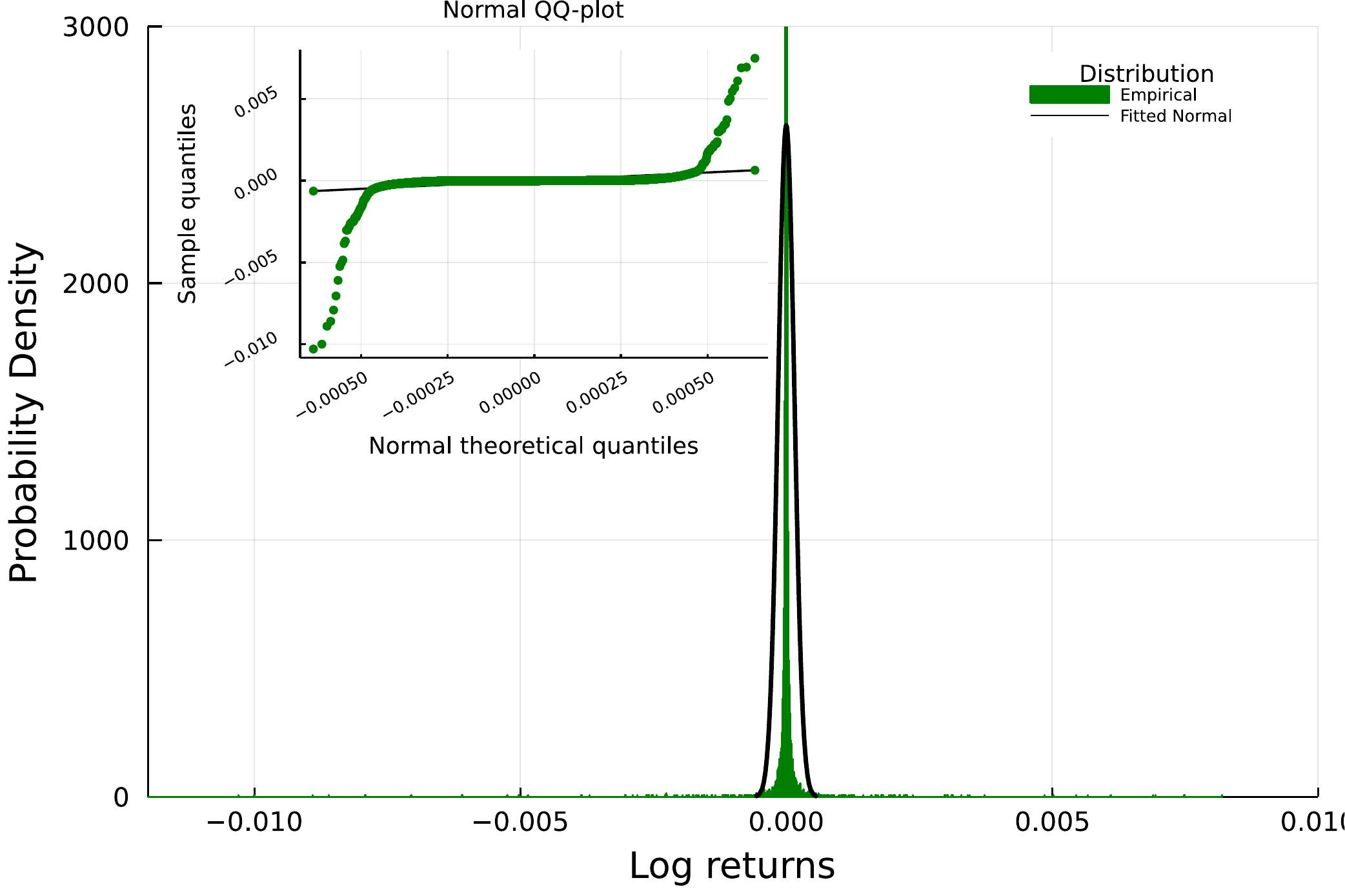}}
    \subfloat[JSE - empirical\label{figb:return distribution}]{\includegraphics[width=.5\textwidth]{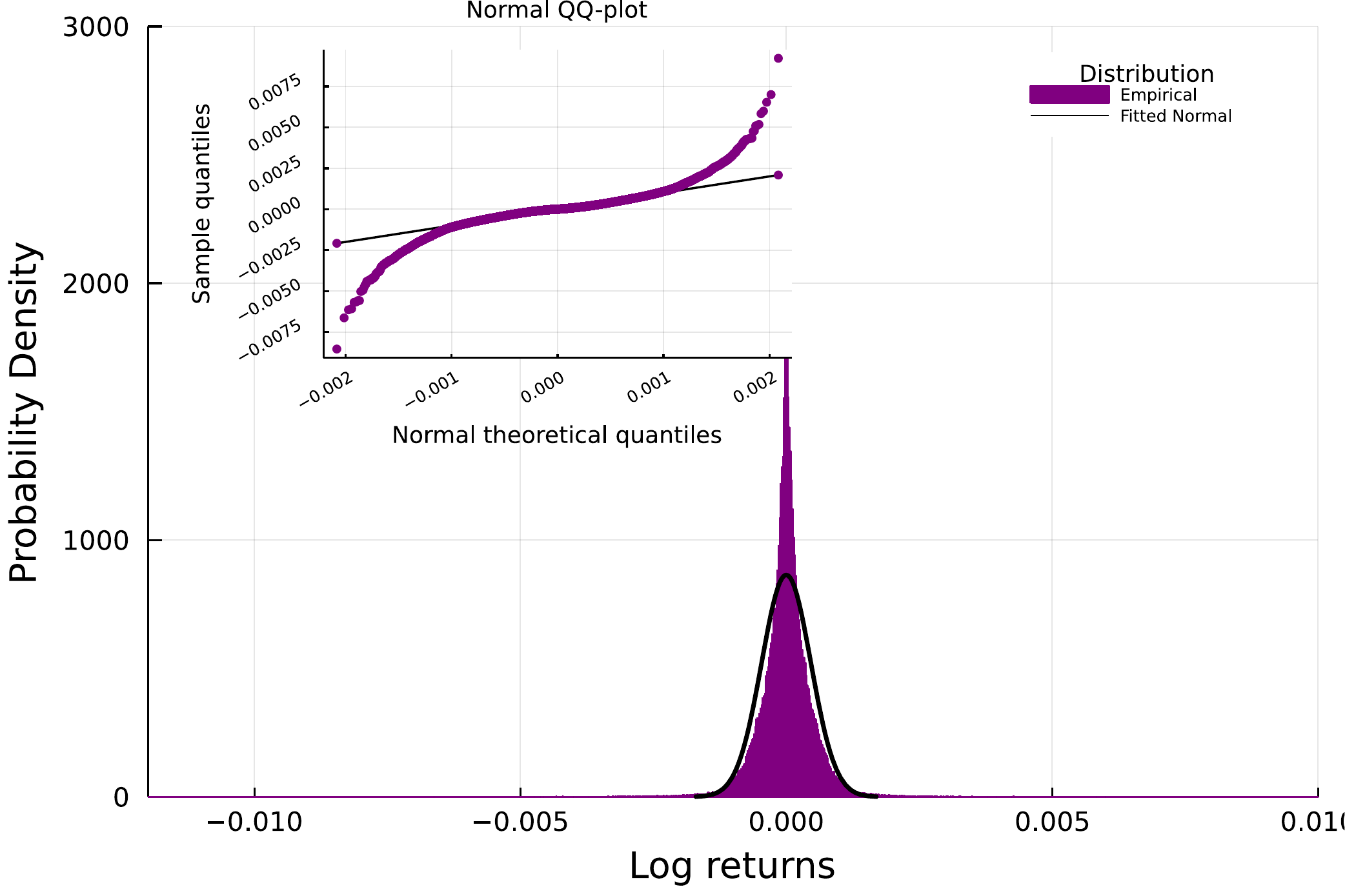}}
    \caption{Tick-by-tick simulated and empirical micro-price log-return distribution properties are depicted along with fitted Normal distributions as references. Simulated price-series were generated by applying the optimal calibration parameters on CoinTossX while empirical results are obtained from JSE level 1 trade-and-quote data. QQ-plots are provided as insets. \label{fig:return distribution}}
\end{figure*}

First, we investigate the distribution of the tick-by-tick returns. \Cref{fig:return distribution} plots the full distribution of the returns with quantile-to-quantile (QQ) plots fitted to a normal distribution provided as insets. Both distributions are highly leptokurtic with heavy tails. Note, however, that empirical returns are more volatile and has its density more evenly distributed away from the mean. This is in contrast to simulated returns which has a larger proportion of its density centred at zero.

\begin{figure*}[!htb]
    \centering
    \subfloat[CoinTossX - calibrated parameters\label{figa:return autocorrelation}]{\includegraphics[width=.5\textwidth]{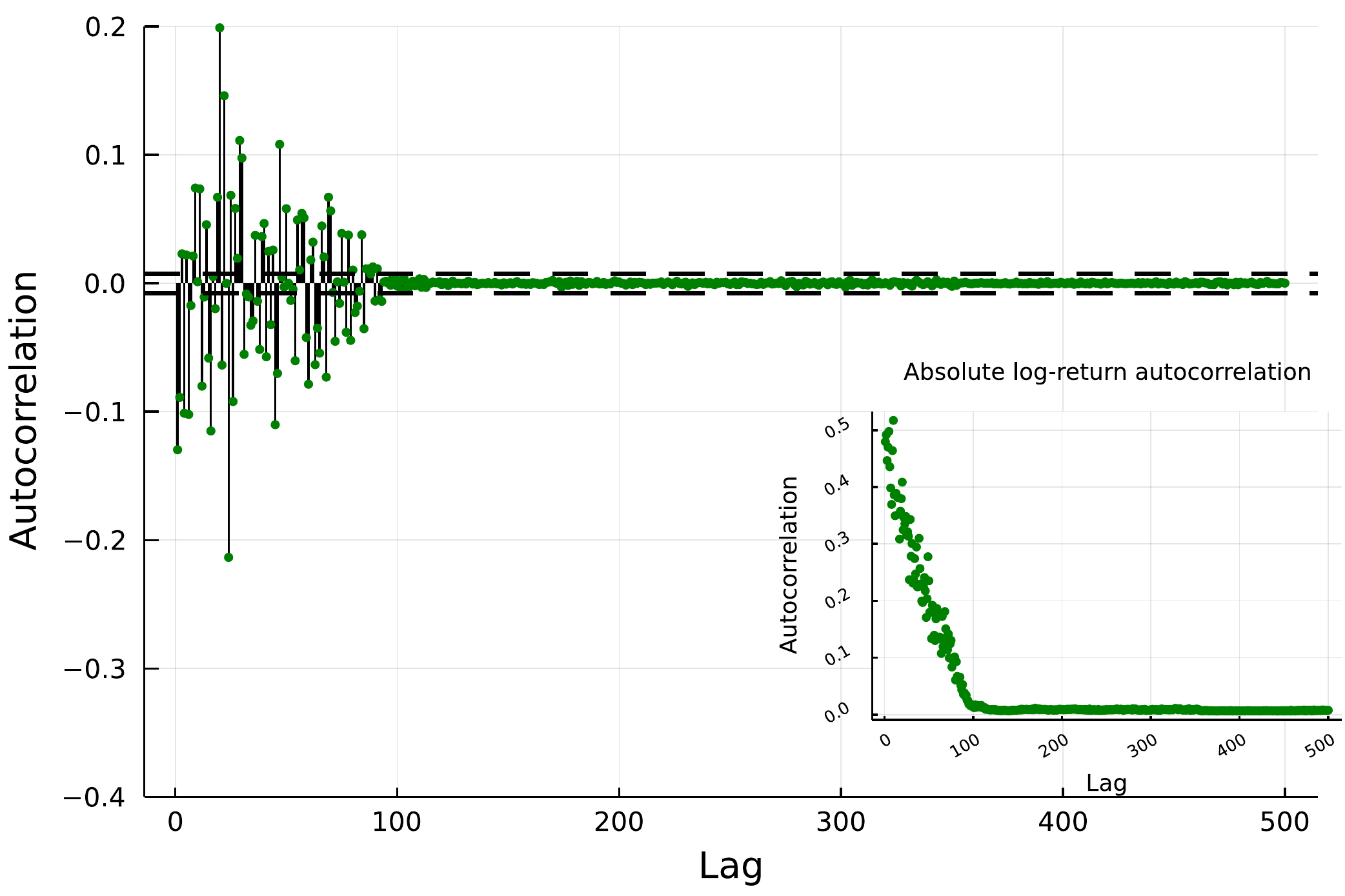}}
    \subfloat[JSE - empirical\label{figb:return autocorrelation}]{\includegraphics[width=.5\textwidth]{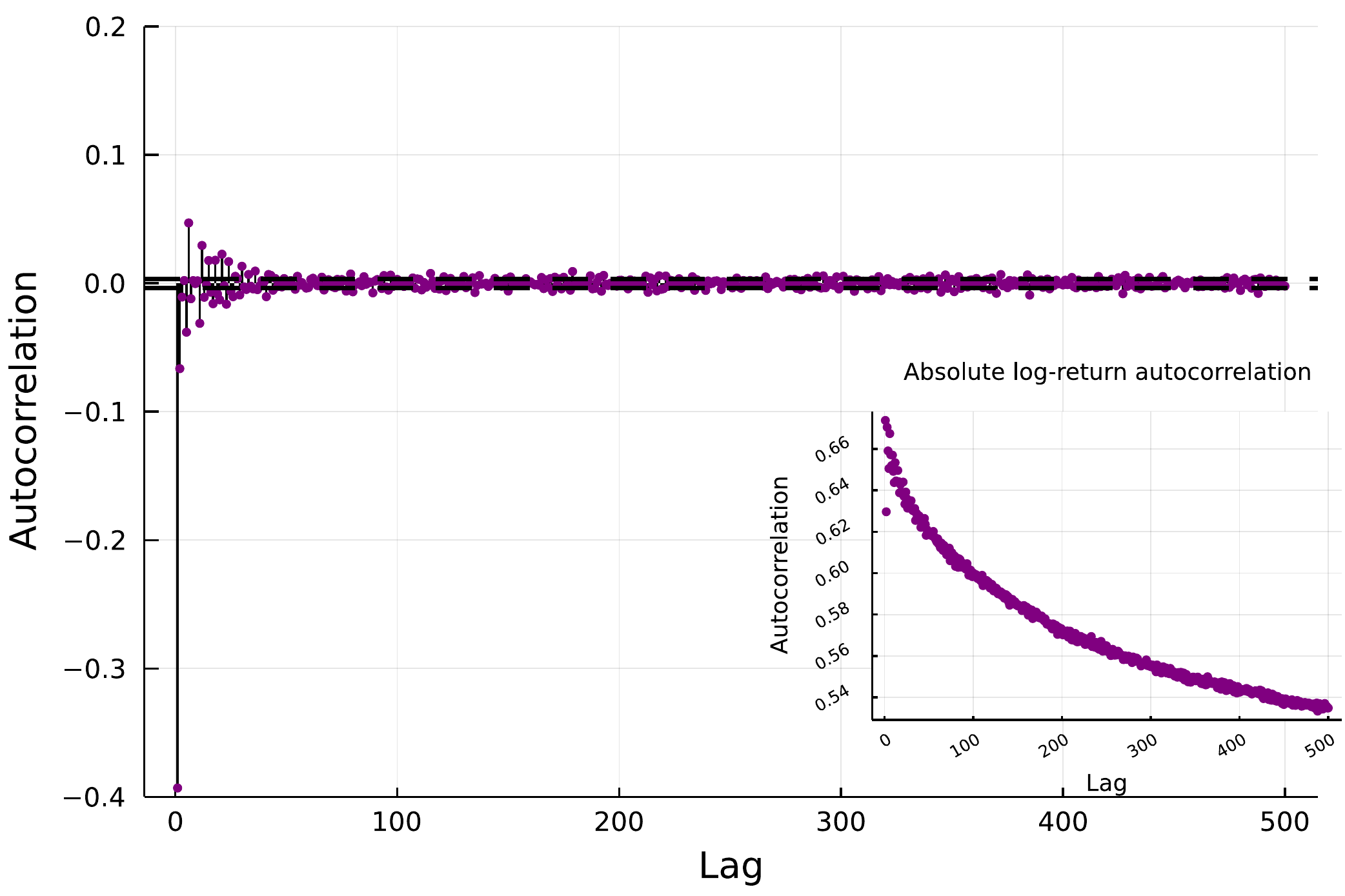}}
    \caption{Tick-by-tick simulated and empirical microprice log-return auto-correlations plots. Simulated price-series were generated by applying the optimal calibration parameters on CoinTossX while empirical results are obtained from JSE level 1 trade-and-quote data. Absolute log-return auto-correlation plots are provided as insets as an indicative measure of long-memory \cite{lillo2004long}. \label{fig:return autocorrelation}}
\end{figure*}

Referring to the log-return auto-correlation plots in \Cref{fig:return autocorrelation}, the strong negative first order auto-correlation seen in empirical returns is indicative of a strong mean-reverting component. This is less prevalent in simulated returns which exhibit significant auto-correlations for the first few lags that decay to zero relatively quickly. We speculate that these differences in long-memory characteristics are due to the significant difference in relative liquidity and order book resilience that are produced by the calibrated parameters.

\begin{figure*}[!htb]
    \centering
    \subfloat[CoinTossX - calibrated parameters\label{figa:order-flow autocorrelation}]{\includegraphics[width=.5\textwidth]{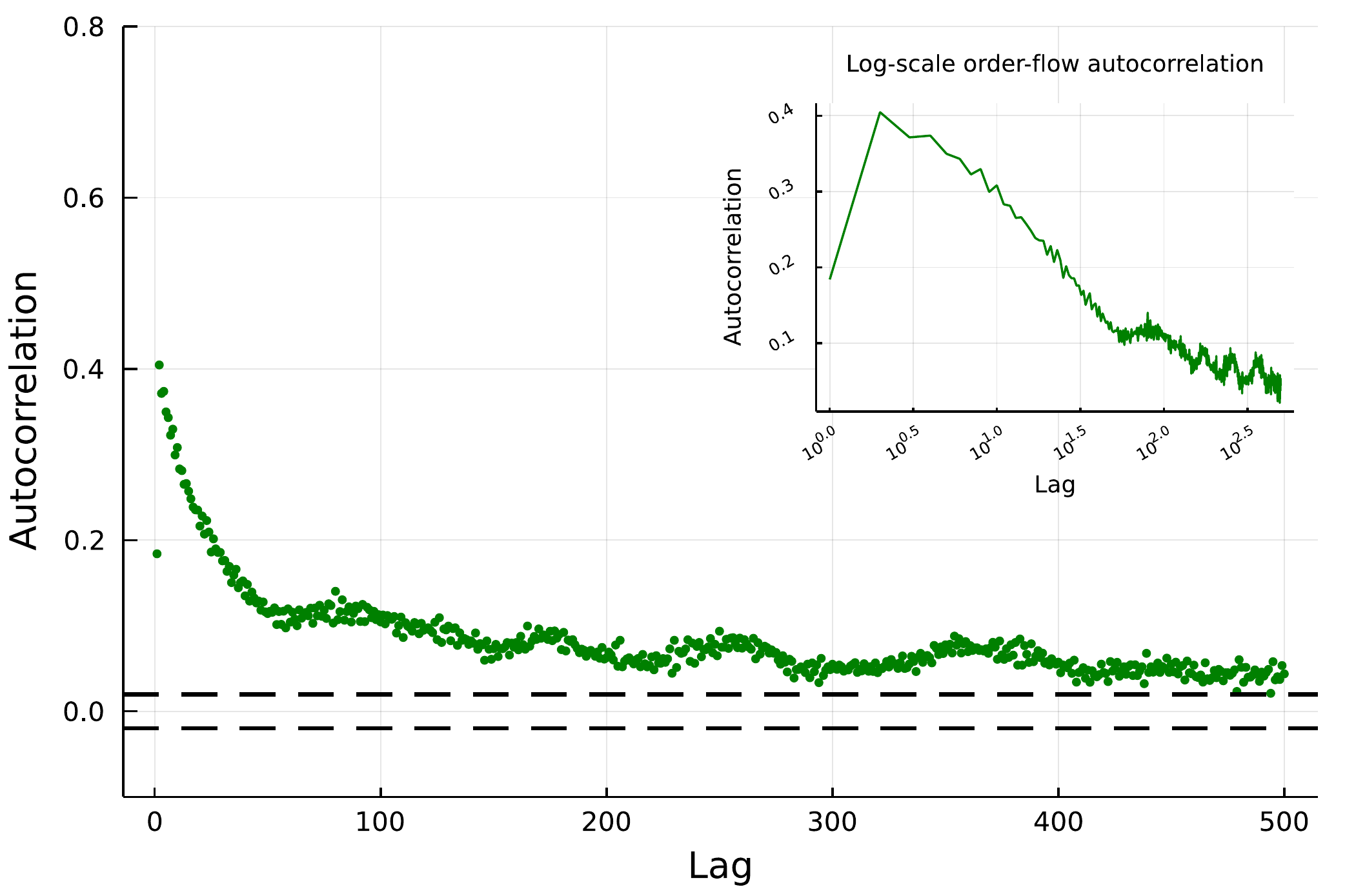}}
    \subfloat[JSE - empirical\label{figb:order-flow autocorrelation}]{\includegraphics[width=.5\textwidth]{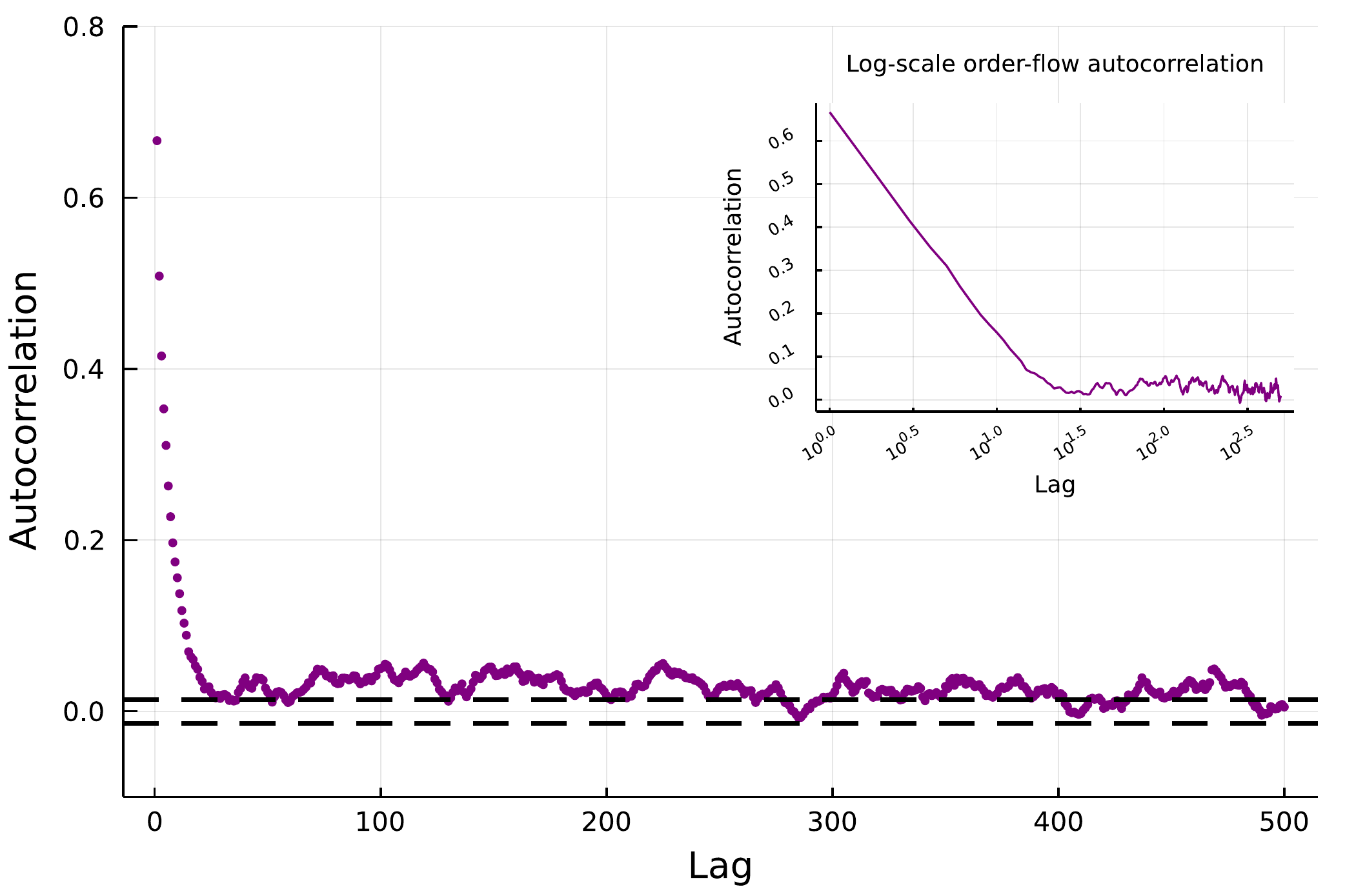}}
    \caption{Order-flow auto-correlation plots based on true and inferred trade signs for simulated and empirical results respectively. Simulated price-series were generated by applying the optimal calibration parameters on CoinTossX while empirical results are obtained from JSE level 1 trade-and-quote data. Provided as insets are the same auto-correlation plots with the lags presented on a $\log_{10}$ scale. \label{fig:order-flow autocorrelation}}
\end{figure*}

The order-flow is represented by a time-series of values with $+1$ for buyer initiated trades, and $-1$ for seller initiated trades. Persistence in the order-flow is then in the sense that buy orders tend to be followed by more buy orders, and sell orders tend to be followed by more sell orders. Possible causes for this are herding behaviours (positive correlation between the behaviour of different investors) and order splitting (positive auto-correlation in the behaviour of single investors). \Cref{fig:order-flow autocorrelation} plots the order-flow auto-correlations from the simulations and empirical data on JSE level 1 trade-and-quote data. The trade classification is performed using the Lee--Ready classification \cite{lee1991inferring} for the empirical data and we compute up to 500 lags between the trades. Furthermore, the same auto-correlation plots presented on a $\log_{10}$ scale are provided as insets. Despite the significant differences in liquidity and the number of trades, we see that both simulated and empirical data sets exhibit persistent order-flows.

\begin{figure*}[!htb]
    \centering
    \subfloat[CoinTossX - calibrated parameters\label{figa:extreme distribution}]{\includegraphics[width=.5\textwidth]{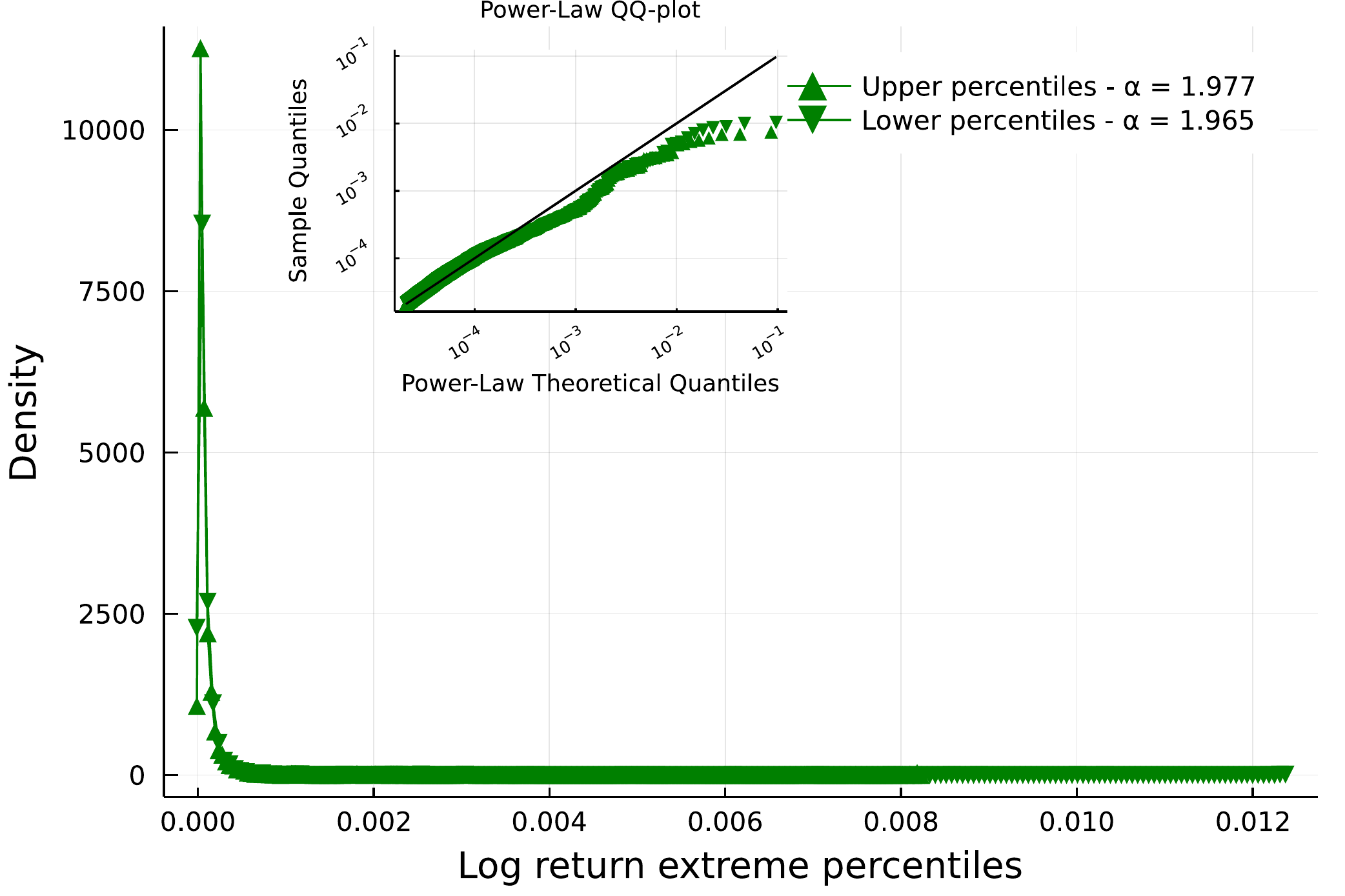}}
    \subfloat[JSE - empirical\label{figb:extreme distribution}]{\includegraphics[width=.5\textwidth]{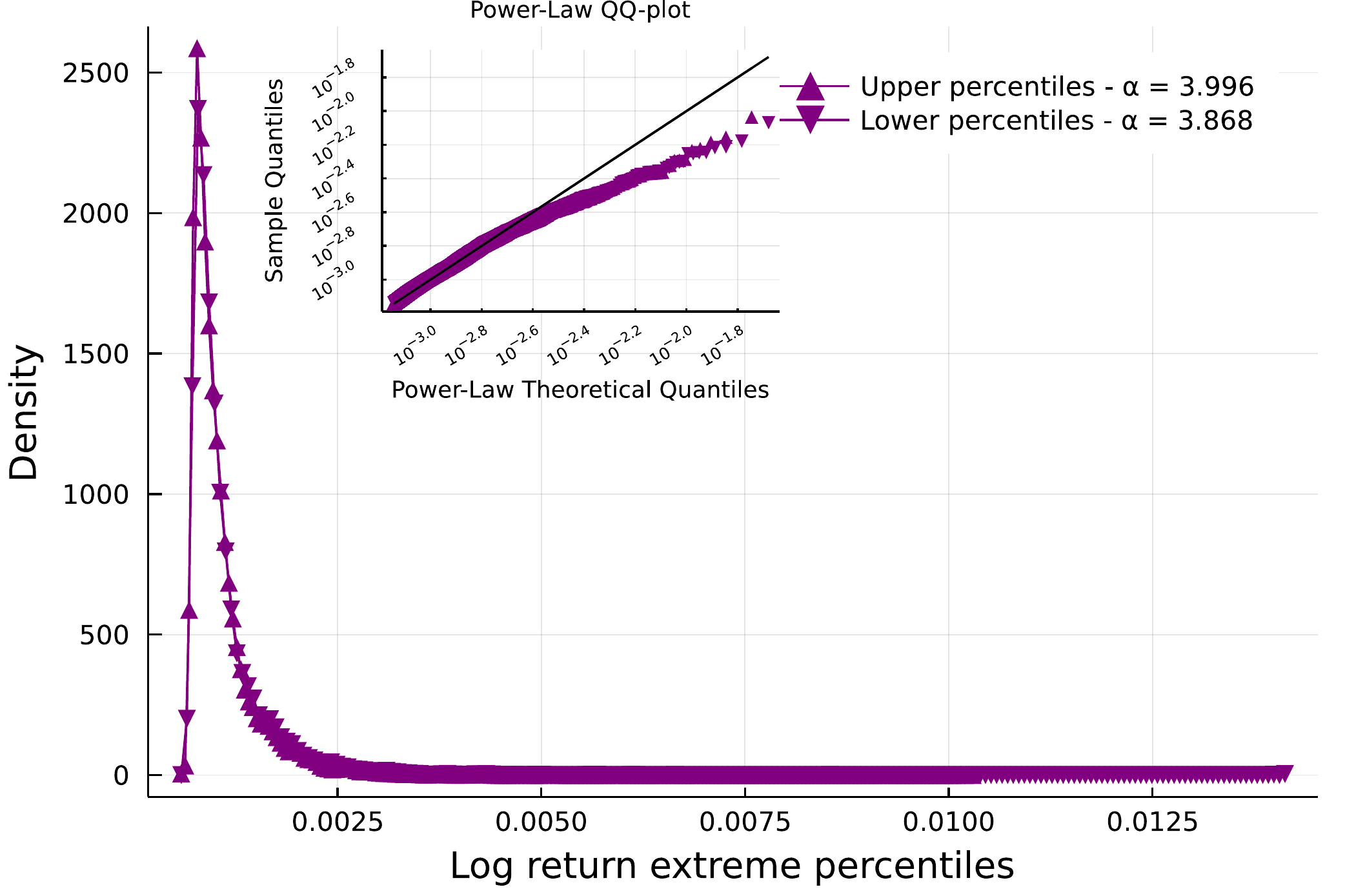}}
    \caption{Distributions of upper and lower extreme log-returns above the 95th percentile and below the 5th percentile respectively computed on tick-by-tick simulated and empirical micro-price log-returns. Simulated price-series were generated by applying the optimal calibration parameter on CoinTossX while empirical results are obtained from JSE level 1 trade-and-quote data. The left and right tails have QQ-plots fitted to a power-law distribution provided as insets. \label{fig:extreme distribution}}
\end{figure*}

Figure \ref{fig:extreme distribution} includes the distributions of the left tail (below the 5th percentile) and the right tail (above the 95th percentile) with QQ-plots fitted to a power-law distribution provided as insets (presented on a log-log scale). The power-law distributions are fitted using maximum likelihood estimation (MLE) where the probability density function is as:
$$
p(x) = \frac{\alpha - 1}{x_{\text{min}}} \left( \frac{x}{x_{\text{min}}} \right)^{-\alpha},
$$
where $x \geq x_{\text{min}} > 0$. The complementary cumulative distribution is:
$$
1 - F(x) = \left( \frac{x}{x_{\text{min}}} \right)^{-\alpha + 1}.
$$
Here we only estimate $\alpha$ because $x_{\text{min}}$ is set as the cutoff percentile. The MLE for $\alpha$ is given as:
$$
\hat{\alpha} = 1+n\left[\sum_{i=1}^{n} \ln \frac{x_{i}}{x_{\min }}\right]^{-1}.
$$
The tail distributions between data sets differ significantly with simulated returns having less of a power-law behaviour than empirical returns as indicated by the smaller estimated tail-index $\alpha$.

Our empirical data only has the level 1 LOB, so the depth profile is only computed on calibrated parameters. To construct these profiles we average the total volume available at the top 7 price levels of the LOB. The resulting curves exhibit a strong decay in the amount of liquidity available at each price level. We also observe asymmetry between the bid and ask mean relative depth profiles. The tendency for most of the volume to be placed at the first level of the LOB is inline with empirical observations \cite{biais1995empirical,gould2013limit} and provides a partial explanation for the power-law price-impact curves.
\begin{figure}[!htb]
    \centering
    \includegraphics[width=.5\textwidth]{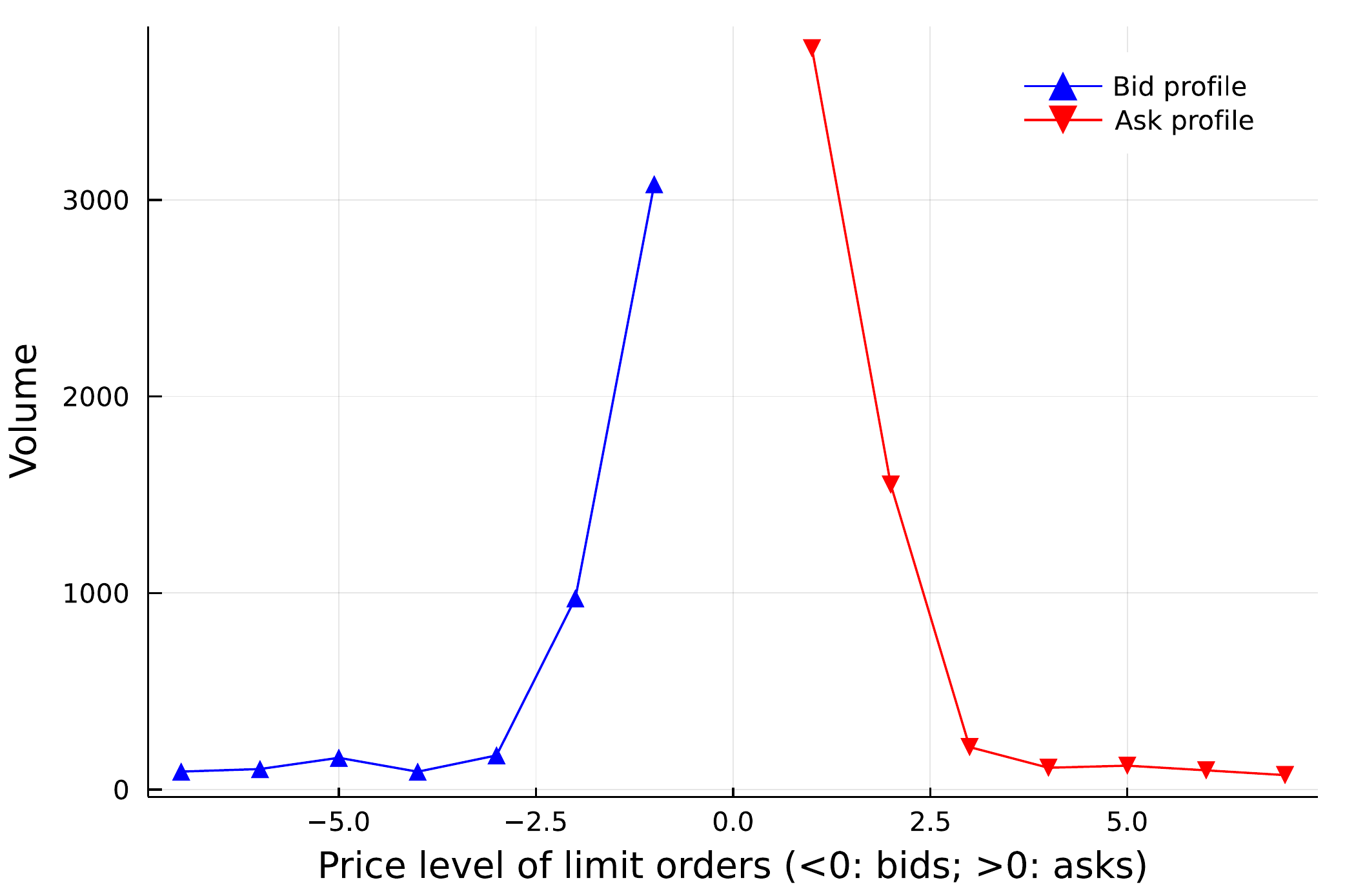}
    \caption{LOB depth profiles on simulated results from optimal calibration parameters as depicted by the average volume at each of the top seven price levels through time. \label{fig:depth profile}}
\end{figure}

\subsection{Price response and impact \label{sec:price impact}}
Here we compare the immediate price impact functions, which are the immediate price responses that are due to transactions, between the empirical and simulated data. Price impact quantifies how a transaction of a given volume affects the best bid/ask price \cite{gebbie2017deviations,gebbiechangjericevich2020comparing}. Let $m_t$ be the mid-price, then we define the impact or immediate price increment of a transaction occurring at time $t_k$ as:
$$
\Delta p_{t_{k}} = \ln\left( m_{t_{k+1}} \right) - \ln\left( m_{t_{k}} \right),
$$
where $t_{k}$ is the time of mid-price before the transaction and $t_{k+1}$ is the time of mid-price immediately after the transaction.\footnote{We make the strict assumption to only use the mid-price immediately after the transaction. For the JSE data set, all transactions are immediately followed by an updated quote. For CoinTossX, we also update the quote immediately following a transaction.}

Additionally, we follow \citet{gebbie2017deviations} and \citet{gebbiechangjericevich2020comparing} and normalise the trade volume as:
$$
\omega_{i j}=\frac{v_{i j}}{\sum_{k=1}^{T_{j}} v_{k j}}\left[\frac{\sum_{j=1}^{M} T_{j}}{M}\right],
$$
where $\omega_{ij}$ is the normalised daily-normalised volume for trade $i$ on day $j$, $T_j$ is the number of trading events on the $j$th day and $M$ is the total number of days. The relationship between price change and transaction size is investigated separately for buyer- and seller-initiated transactions. The transactions in the JSE data set are classified according to the Lee--Ready rule \cite{lee1991inferring}.

\begin{figure*}[!htb]
    \centering
    \subfloat[CoinTossX - calibrated parameters\label{figa:price impact}]{\includegraphics[width=.5\textwidth]{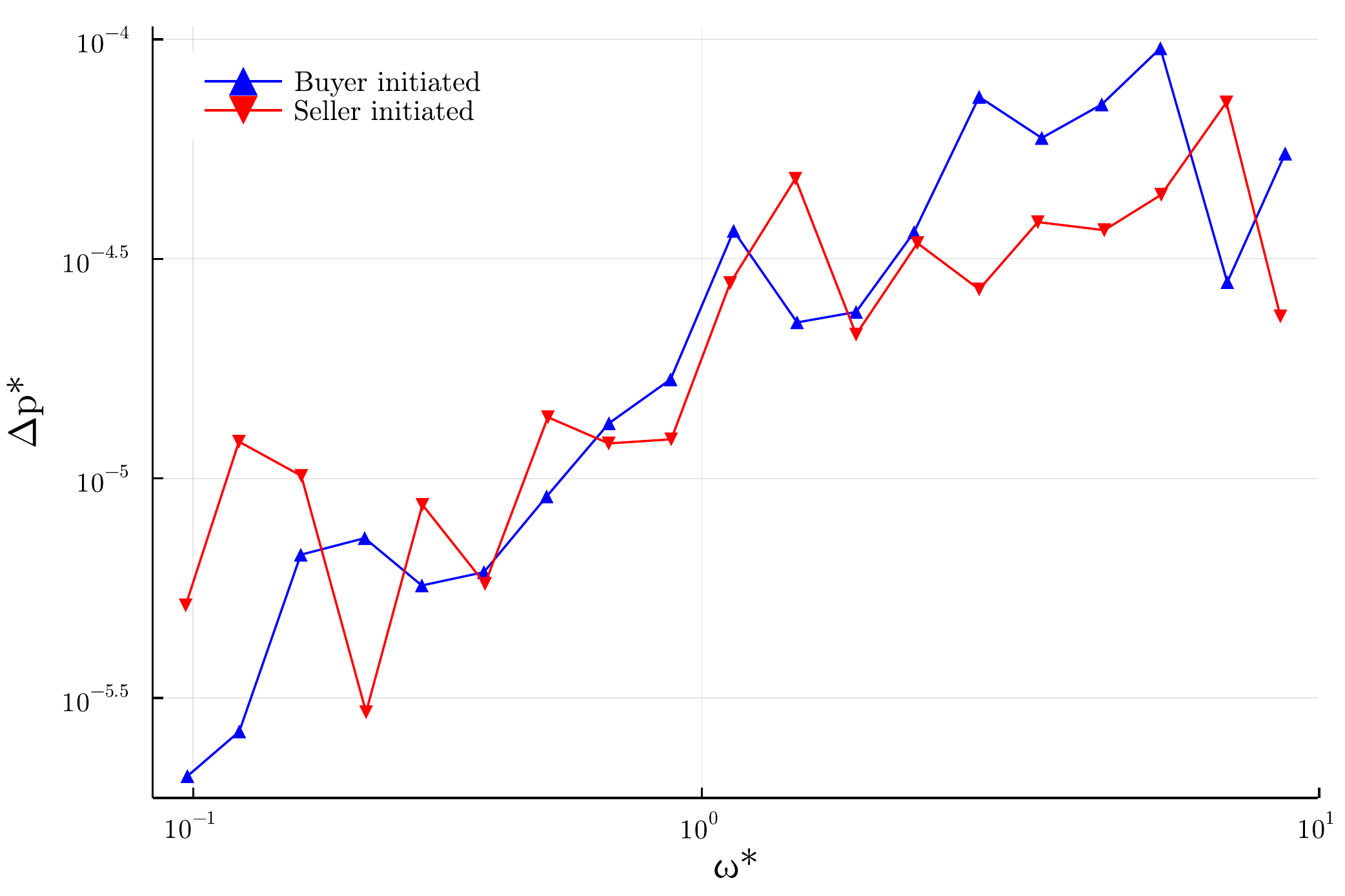}}
    \subfloat[JSE - empirical\label{figb:price impact}]{\includegraphics[width=.5\textwidth]{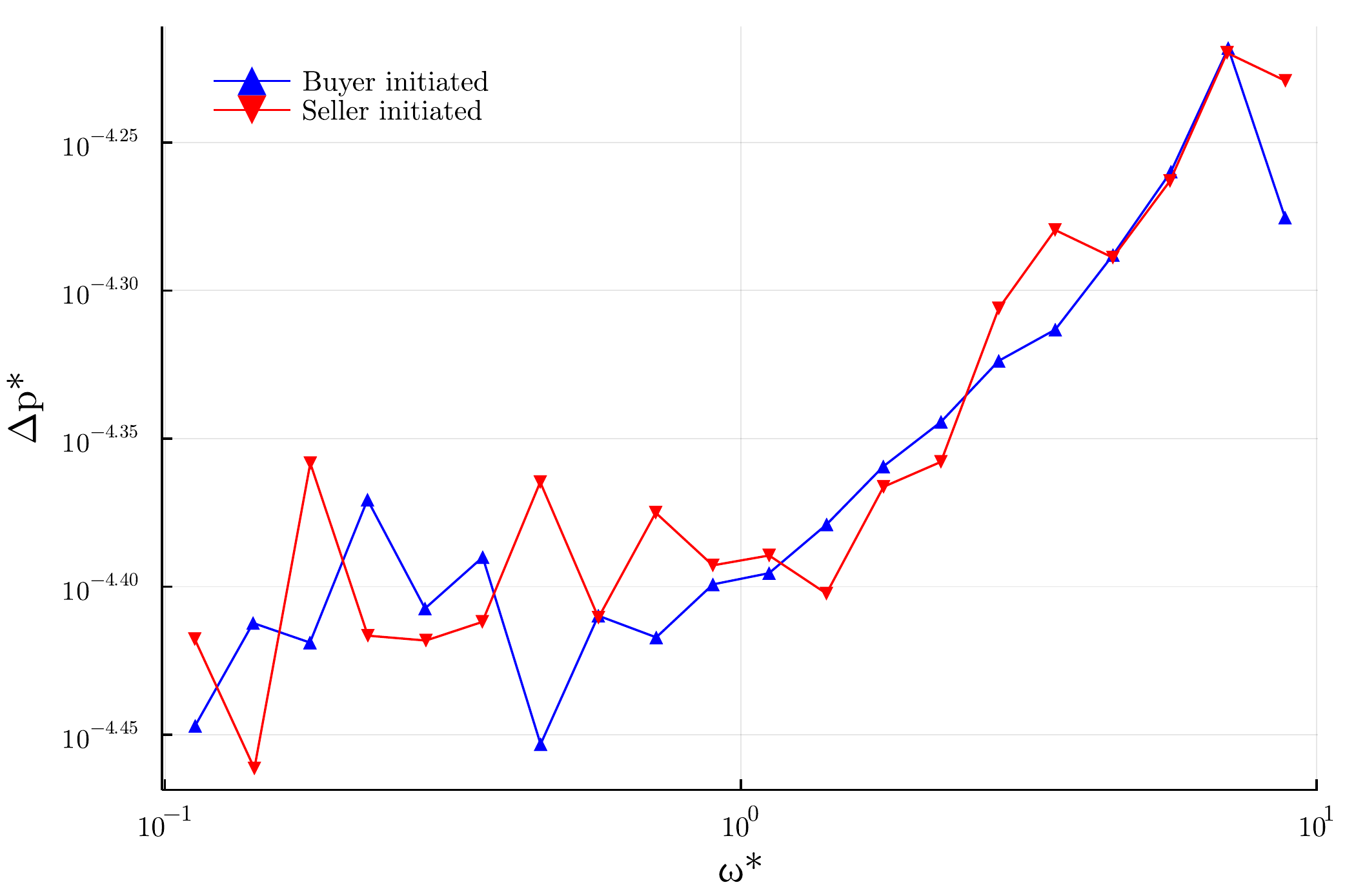}}
    \caption{Individual buyer-initiated and seller-initiated price impact curves plotted on a log-log scale with average daily-normalised volumes $\omega^*$ on the x-axis and average price increments $\Delta p^*$ on the y-axis. Simulated price-series were generated by applying the optimal calibration parameters on CoinTossX while empirical results were obtained from JSE level 1 trade-and-quote data.\label{fig:price impact}}
\end{figure*}

In \Cref{fig:price impact} we investigate the price impact for a given volume size. We create 20 logarithmically spaced normalised volume bins between $[10^{-1}, 10^{1}]$. For each of these volume bins we compute the average price change $\Delta p^*$ and the average normalised volume $\omega^*$ over the entire data set. The relationship between the average price change $\Delta p^*$ and the average volume bin $\omega^*$ is then plotted on a log-log scale for buyer- and seller-initiated transactions. Consistent with previous findings \cite{gebbie2017deviations,gebbiechangjericevich2020comparing,LFM2003}, we see that there is a power-law relationship (linear on a log-log scale) between the price change and normalised volume in both the simulated and empirical data.

\section{Discussion} \label{sec:discussion}
\subsection{Limitations}

For the purpose of future improvements in market simulations with CoinTossX we mention two key shortcomings of the model in this work. The first and most important for dynamics and agent activation being that of the limit price placement.

As shown by the depth profiles, limit price placements are sampled from a Gamma distribution which will have most of their density at or just below the best bid/ask. This results in most of the volume being placed at the best bid or ask price level. Although this loosely conforms with empirical observations, this characteristic tends to result in little to no liquidity taker activation in the presence of too many liquidity providers because trade volumes were often insufficient to break through the first level. This can then result in very little price movement and thus little variation in log-returns. 

This shortcoming is further aggravated, but not caused, by the choice of a power-law volume distribution. There is a tension because too many liquidity takers will mean too little liquidity even at level 1. Deeper levels in the order book are shallow with most volume directed at level 1. Clusters of trades or large individual trades breaking through the first level would subsequently break through all deeper levels and thereby emptying one side of the order book. With one side empty, subsequent limit prices placements sampled through the Gamma distribution would tend to place prices far from the contra best --- resulting in a large spread. This behaviour would repeat as large price movements can increase the aggression of liquidity takers. The end result being unrealistically large price changes, sometimes into the negatives. Additionally, once large price changes occur, the fail-safe preventing volatility-auctions to trigger would prevent LT agents submitting further market orders.

The second feature that may have had adverse effects on simulation dynamics were agent activation rules. In particular, fundamentalists were assigned a fixed value perception for the day. If the perceived value was far from the starting mid-price, consistent upward/downward trends would persist throughout the day with little market pressure being applied to the opposite side. This may give evidence for the existence of contrarian strategies in real markets.

\subsection{Future extensions and reactive agents}

Having an independent matching infrastructure at our disposal means that greater attention can be applied to the model specification, configurations and attributes as the mechanics of order matching is externally provided as a given. To implement a truly asynchronous actor-based model, the modelling framework needs to be able to react to streaming data in an event based framework. This will require the model management system and its integration with the data feed to become reactive. This is how real financial markets function. This will require a reactive framework such as \href{https://akka.io/}{akka} to implement a truly multi-agent model framework (as one example) or the extension of the existing code base.\footnote{The requirement for reactive agents can be implemented using our Julia components by a simple migration into a framework such as \href{https://github.com/JuliaGizmos/Reactive.jl}{Reactive.jl} or \href{https://github.com/biaslab/Rocket.jl}{Rocket.jl} to replace our hybrid implementation.}

Doing so will allow the system to include additional classes of agents, {\it e.g.,} RL agents, learning agents, distressed agents and multi-asset agents that can potentially induce cross asset correlations. Each asset can then still be traded by at least the 3 basic agent formulations: the two LF and HF agents who will then be operating in event time. Additional agent classes can then inject trades into the matching engine and the various model management systems can then monitor the order book and trade events in order to make trading decisions. 

Examples of multi-asset agent-based models can be found in \cite{bohm2005mean, ponta2011multi}. These models have historically been based on sequential order submission and market clearing. However, with a bit of work these can be extended to become compatible with an event driven asynchronous framework moderated by a matching engine. 

\section{Conclusion \label{sec:Conclusion}}

We have presented an agent-based model in an artificial exchange framework that is able to recover price-impact while replicating the conventional stylised facts by directly interacting with a decoupled Matching Engine (ME) that manages orders away from the software relating to agent interactions and rules. 

The key innovation here is that the model updates no longer occur in chronological or calendar time. In many models such as that of \citet{leal2016rock}, the activations are random, but the model takes place in calendar time with sequential market clearing. This means that the agent rules do not depend on market clearing prices and additional constraints are needed to create realism.

Here the Model Management System (MMS) polls the market data feed based on asynchronous model time, rather than be entirely driven by the event time of the market. This is a ``hybrid framework" where we have not isolated the MMS from the ME entirely. This was a convenient choice made to reduce compute times for the work done here. However, the system was designed and setup for the ``reactive framework'' which is what future work should focus on. 

We recall that the MMS plays the same roles as the OMS/EMS in a typical trading technology stack. The hybrid solution is computationally convenient here because we make use of the model specification that the only trade events experienced by the matching engine will be those triggered and injected from the MMS environment. This means that we do not need to implement a truly reactive framework as there are no other agents that can trigger events for which the agent-based model needs to react to, {\it i.e.,} there are no agents outside of those from the model specification. 

The complexity of balancing a reactive agent-based modelling system with the convenience of fast calibration is central to the design decisions taken here. The use of Nelder-Mead (NM) with the heuristic Threshold-Acceptance (TA) approach was similarly pragmatic in as much as we have opted for a fast calibration methodology that will allow us to quickly get to indicative parameter values. Hence a practical framework that allows one to quickly get indicative parameters and generally replicate market stylised facts with relative effectiveness. The framework balances theoretical and statistical rigour, computational convenience and implementation realism. We also made the choice of having more moments than there are calibration parameters; here 5 parameters but 9 moments. 

The agent-based model's optimal calibration parameters were found to provide more consistent time-series behaviour across seeds than parameter values selected based on the sensitivity analysis. It was noted that simulated prices obtained from sensitivity analysis parameters would occasionally either trigger the volatility-auction fail-safe or close out prices until they remained unchanged --- thereby preventing any further actions from agents who depend on price movements. 

In the instance where the MMS is one of many such frameworks connected to the market then the MMS needs to listen for externally triggered event types that are not the result of its own agent responses. This is computationally expensive. This is important if one is then going to implement alternative agents exogenous to those specified in this particular MMS, {\it e.g.,} distressed agents, learning agents, a specific implementation of a particular trading algorithm, or particular types of agent rules that one would like to investigate relative to the core framework. The system as it is currently designed and configured can accommodate this. However, this was not implemented here due to computational efficiency requirements and is left to future work.

\section*{Reproducibility}
The specific version of CoinTossX used can be found at: \href{https://github.com/IvanJericevich/CoinTossX/tree/abm-experiment}{https://github.com/IvanJericevich/CoinTossX/tree/abm-experiment}. Code and instructions for replication can be found at: \href{https://github.com/IvanJericevich/IJPCTG-ABMCoinTossX}{https://github.com/IvanJericevich/IJPCTG-ABMCoinTossX}. 
\section*{Acknowledgements \label{sec:Acknowledgements}}
We would like to thank Dharmesh Sing for extensive support with regards to CoinTossX.

\balance
\bibliographystyle{elsarticle-harv}
\bibliography{ABMMS-CoinTossX}

\onecolumn
\appendix
\section{Market data and messaging}\label{app:data}

The basic procedure for submitting orders to CoinTossX is demonstrated in pseudocode \ref{algo:injectsimulation} of \ref{ssec:pseudo}. We will discuss how different order type submissions are specified. Instead of specifying limit order expiry durations, we cancel orders manually once their active time limit is reached. Cancellations require an order identification number, a side and a price. These fields should match the limit order to be cancelled and is determined by the decision rules of the liquidity providers. Limit orders require a volume, side and price which are similarly determined by liquidity providers. Market order messages only require a side and volume. These fields are determined by the decision rules of the liquidity takers.

We deviate slightly from original CoinTossX architecture in order to receive the market data from CoinTossX. Since Julia currently cannot integrate with Aeron media driver, we open another port in the matching engine for the forwarding of byte encoded messages via UDP. This is in response to issues regarding the LOB snapshot functionality previously used in \citet{arxiv2021hawkes}. First, requesting a snapshot of the LOB is not realistic from a trading/exchange point of view. Second, in the event of a large build of of orders in the LOB, requesting it often would compromise latency. Therefore, we assign a dedicated listener to the modelling framework in Julia that will continuously wait and receive market data messages as they occur. The LOB is built with each message update as a separate process and will form the ``lit order book''.

In terms of the feedback from each order submission, the market data feed is easy to customise and currently adopts the following format. Each report is represented by a single string which is encoded into a vector of bytes and sent via UDP to a specified port and address. The message is decoded back into a string in Julia to be processed. The reason for adopting this process is so that the transmitting speed is as fast as possible. This method also automates the process of receiving market data as opposed to manually requesting execution reports.

The snippet below provides examples of messages transmitted. The first field provides the number of milliseconds since the unix epoch at which the event(s) arrived. The second specifies the event type which will either be: ``New'' (limit order), `Trade'' or ``Cancelled''. The third field specifies the side of the order. The fourth field provides the agent's identity. More importantly, the last field is given in vector form where each sub-vector contains the ID, price and volume of the order. In the second last case where a trade walked the LOB and executed against the first two limit orders, two different sub-vectors are recorded and corresponds to the single market order that was split into two and executed at two different prices.
\begin{verbatim}
1621159225272,New,Buy,HF|1,50,100
1621159230899,New,Buy,HF|2,49,100
1621159235548,Trade,Sell,LF|1,50,100|2,49,50
1621159242341,Cancelled,Buy,HF|2,0,0
\end{verbatim}

The data output follows the format below and is only written out to file once all clients are logged out at the end of the simulation so that latency is not compromised during trading sessions. Execution times are printed at millisecond precision. Each order is associated with the ID of the trader who submitted the order (``TraderMnemonic''), the ID assigned to the order (``ClientOrderId'' used to track cancelled and executed orders) as well as the price, volume, side and type of the order.\footnote{Note that a single ``client'' logged in to CoinTossX need not be associated with a single trader. The ``clients'' in CoinTossX may be interpreted as a group of traders or an investment firm.} Orders specified by the ``New'' type are limit orders. As shown below, CoinTossX duplicates individual trade events by printing both the full ``unbroken'' market order submitted as well as the split trades that have been executed at multiple prices. Unbroken market orders have a zero execution price. The unbroken market orders retain their assigned order IDs but the ``split'' market orders are assigned order IDs corresponding to the limit orders against which they were executed. Similarly, cancel orders are assigned the ID of the corresponding limit orders.
\begin{verbatim}
"DateTime","TraderMnemonic","ClientOrderId","Price","Volume","Side","Type"
"2021-05-19 11:35:34.827","1","1","20","100","Buy","New"
"2021-05-19 11:35:43.405","1","3","0","150","Sell","Trade"
"2021-05-19 11:35:43.406","1","2","49","100","Sell","Trade"
"2021-05-19 11:35:43.415","1","1","20","50","Sell","Trade"
"2021-05-19 11:35:47.027","1","2","49","0","Buy","Cancelled"
"2021-05-19 11:35:59.973","1","4","50","100","Buy","New"
\end{verbatim}

\newpage
\section{Simulation State-Flow} \label{ssec:pseudo}

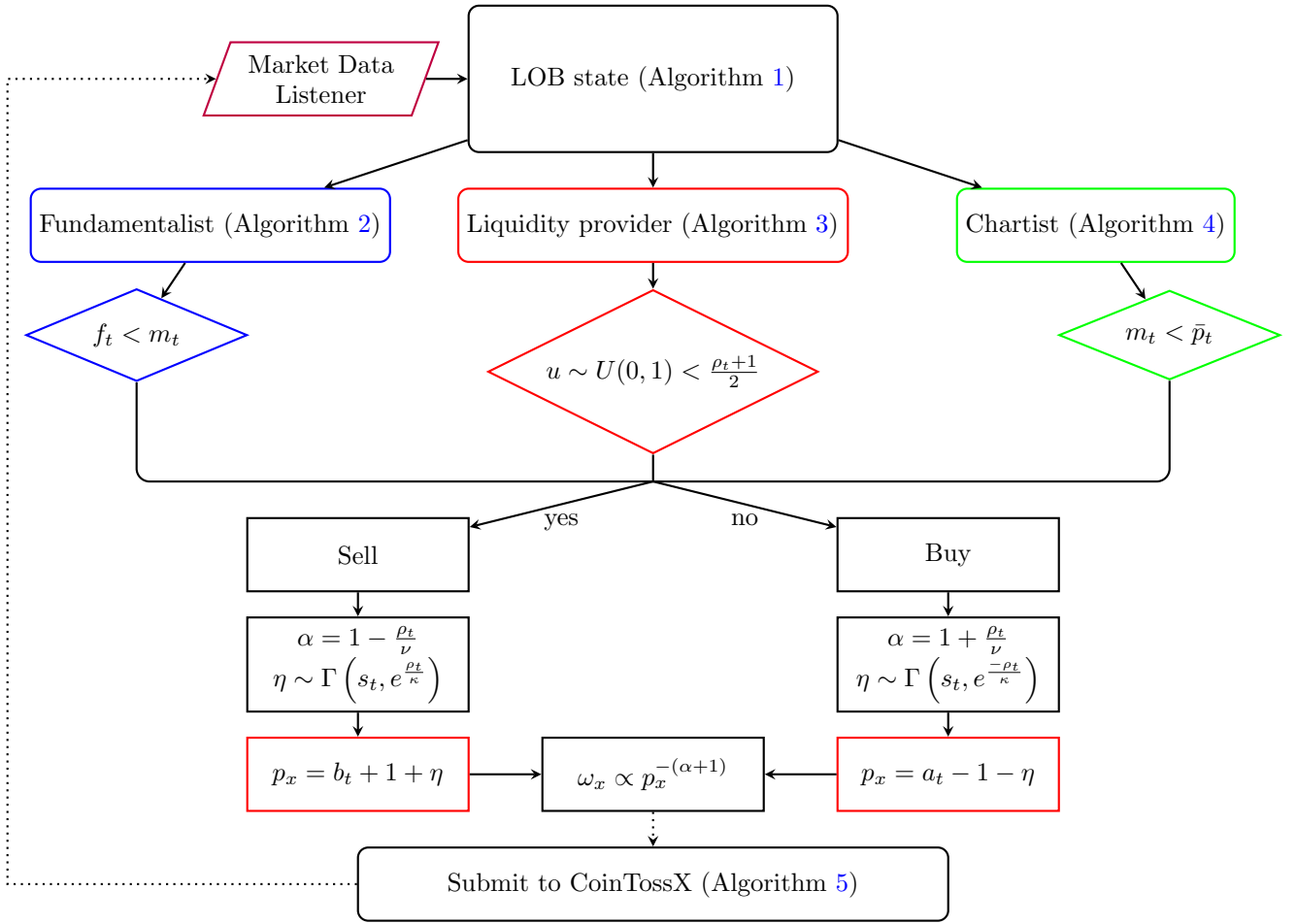
\begin{figure}[H]
    \centering
    \begin{tikzpicture}[node distance=2cm, scale = 0.25]
        \node (vi) [rectangle,rounded corners,minimum width=3cm,minimum height=1cm,text centered,draw=blue,thick] {Fundamentalist (Algorithm \ref{algo:fundamentalists})};
        \node (lp) [rectangle,rounded corners,minimum width=3cm,minimum height=1cm,text centered,draw=red,thick,right of=vi,xshift=4cm] {Liquidity provider (Algorithm \ref{algo:liquidtyprovider})};
        \node (tf) [rectangle,rounded corners,minimum width=3cm,minimum height=1cm,text centered,draw=green,thick,right of=lp,xshift=4cm] {Chartist (Algorithm \ref{algo:chartists})};
        \node (lob) [rectangle,rounded corners,minimum width=5cm,minimum height=2cm,text centered,draw=black,thick,above of=lp] {LOB state (Algorithm \ref{algo:lob state})};
        \node (lpside) [diamond,aspect=2,minimum width=3cm,minimum height=1cm,text centered,draw=red,thick,below of=lp,align=center] {$u \sim U(0, 1)< \frac{\rho_t + 1}{2}$};
        \node (viside) [diamond,aspect=2,minimum width=3cm,minimum height=1cm,text centered,draw=blue,thick,below of=vi,align=center,xshift=-1cm,yshift=0.5cm] {$f_t < m_t$};
        \node (tfside) [diamond,aspect=2,minimum width=3cm,minimum height=1cm,text centered,draw=green,thick,below of=tf,align=center,xshift=1cm,yshift=0.5cm] {$m_t < \bar{p}_t$};
        \coordinate[below of=lpside,yshift=0.5cm] (vertex);
        \node (buy) [rectangle,minimum width=3cm,minimum height=1cm,text centered,draw=black,thick,right of=lpside,align=center,yshift=-2.5cm,xshift=2cm] {Buy};
        \node (sell) [rectangle,minimum width=3cm,minimum height=1cm,text centered,draw=black,thick,left of=lpside,align=center,yshift=-2.5cm,xshift=-2cm] {Sell};
        \node (alphabuy) [rectangle,minimum width=3cm,minimum height=1cm,text centered,draw=black,thick,below of=buy,align=center,yshift=0.5cm] {$\alpha = 1 + \frac{\rho_t}{\nu}$\\$\eta \sim \Gamma \left(s_t, e^{\frac{-\rho_t}{\kappa}} \right)$};
        \node (alphasell) [rectangle,minimum width=3cm,minimum height=1cm,text centered,draw=black,thick,below of=sell,align=center,yshift=0.5cm] {$\alpha = 1 - \frac{\rho_t}{\nu}$\\$\eta \sim \Gamma \left(s_t, e^{\frac{\rho_t}{\kappa}} \right)$};
        \node (pricebuy) [rectangle,minimum width=3cm,minimum height=1cm,text centered,draw=red,thick,below of=alphabuy,align=center,yshift=0.5cm] {$p_x = a_t - 1 - \eta$};
        \node (pricesell) [rectangle,minimum width=3cm,minimum height=1cm,text centered,draw=red,thick,below of=alphasell,align=center,yshift=0.5cm] {$p_x = b_t + 1 + \eta$};
        \node (volume) [rectangle,minimum width=3cm,minimum height=1cm,text centered,draw=black,thick,below of=vertex,align=center,yshift=-2cm] {$\omega_x \propto p_x^{-(\alpha+1)}$};
        \node (cointossx) [rectangle,rounded corners,minimum width=8cm,minimum height=1cm,text centered,draw=black,thick,below of=volume,align=center,yshift=0.5cm] {Submit to CoinTossX (Algorithm \ref{algo:injectsimulation})};
        \node[align=center,xshift=-2.5cm,trapezium,trapezium left angle=70,trapezium right angle=110,minimum width=3cm,minimum height=1cm,text centered,draw=purple,thick,left of=lob] (listener) {Market Data\\Listener};
        \draw [thick,->,>=stealth,dotted] (cointossx) -| (-11, 8) -- (listener);
        \draw [thick,->,>=stealth] (listener) -- (lob);
        \draw [thick,->,>=stealth] (lob) -- (lp); \draw [thick,->,>=stealth] (lob) -- (vi); \draw [thick,->,>=stealth] (lob) -- (tf);
        \draw [thick,->,>=stealth] (lp) -- (lpside); \draw [thick,->,>=stealth] (vi) -- (viside); \draw [thick,->,>=stealth] (tf) -- (tfside);
        \draw [thick,-,>=stealth,rounded corners=5pt] (tfside)|-(vertex); \draw [thick,-,>=stealth,rounded corners=5pt] (viside)|-(vertex); \draw [thick,-,>=stealth] (lpside) -- (vertex);
        \draw [thick,->,>=stealth] (vertex) -- node[anchor=north] {no} (buy); \draw [thick,->,>=stealth] (vertex) -- node[anchor=north] {yes} (sell);
        \draw [thick,->,>=stealth] (sell) -- (alphasell); \draw [thick,->,>=stealth] (buy) -- (alphabuy);
        \draw [thick,->,>=stealth] (alphasell) -- (pricesell); \draw [thick,->,>=stealth] (alphabuy) -- (pricebuy);
        \draw [thick,->,>=stealth] (pricesell) -- (volume); \draw [thick,->,>=stealth] (pricebuy) -- (volume);
        \draw [thick,->,>=stealth,dotted] (volume) -- (cointossx);
    \end{tikzpicture}
    \caption{ABM simulation state flow diagram \label{fig:abm state flow}. See Table \ref{tab:app:definitions} for the definition of the variables and concepts used. The algorithm that updates the state of the Limit-Order Book (LOB) is given in Algorithm \ref{algo:lob state}. The agent activation algorithms are given by algorithms \ref{algo:fundamentalists}, \ref{algo:liquidtyprovider}, and \ref{algo:chartists} for the implementations for the Fundamentalist, Liquidity provider and Chartist agents, respectively. The order injection algorithm with the order submission logic is given in Algorithm \ref{algo:injectsimulation}.}
\end{figure}

Here we provide the simulation pseudocode for the agent-base model simulation framework. The framework starts with the basic logic for the updating the Limit-Order Book (LOB) (Algorithm \ref{algo:lob state}), agent can then be activated from one of the three classes of agents: fundamentalists (Algorithm \ref{algo:fundamentalists}), liquidity providers (Algorithm \ref{algo:liquidtyprovider}), and chartists (Algorithm \ref{algo:chartists}), and finally the submission of orders into the Matching Engine (ME) (Algorithm \ref{algo:injectsimulation} to complete the order and agent activation sequence. The MMS listens for activation events. The Matching engine (ME) is separated from the Model Management System (MMS) as shown in Figure \ref{fig:flowchart}). The relationships between the algorithms in the MMS are shown in the simulation state-flow diagram in Figure \ref{fig:abm state flow}. The definitions of key variables and parameters used in the simulation state-flow are given in Table \ref{tab:app:definitions}. 

\begin{table}[ht]
\begin{tabular}{p{2.75cm} p{14.5cm}}
\toprule
Definition & Description \\
\toprule
Order & An order is a tuple \( x = (p_x, \omega_x, t_x) \) submitted at time \( t_x \) with price \( p_x \) and size \( \omega_x > 0 \) and is a commitment to buy (sell) up to \( \omega_x \) units of a traded asset at a price no greater (less) than \( p_x \). \\
\midrule
Limit order book & A Limit Order-Book (LOB) \( \mathcal{L}(t) \) is the set of all active orders in a market at time \( t \). The active orders in a LOB can be partitioned into the set of active buy orders \( \mathcal{B}(t) \) and the set of active sell orders \( \mathcal{A}(t) \). The LOB can then be considered as a set of queues at specified prices. \\
\midrule
Depth & The bid-side depth and ask-side depth at price \( p \) and time \( t \) are given by: \( n^b(p, t) := \sum_{x \in \mathcal{B}(t)|p_x = p} \omega_x \) and \( n^a(p, t) := \sum_{x \in \mathcal{A}(t)|p_x = p} \omega_x \). \\
\midrule
Spread & The bid-ask spread at time \( t \) is \( s(t) := a(t) - b(t) \). \\
\midrule
Mid-price & The mid-price at time \( t \) is \( m(t) := [a(t) + b(t)] / 2 \). \\
\midrule
Imbalance & The order imbalance at time \( t \) is \( \rho(t) := \frac{\sum_{x \in \mathcal{B}(t)} \omega_x - \sum_{x \in \mathcal{A}(t)} \omega_x}{\sum_{x \in \mathcal{B}(t)} \omega_x + \sum_{x \in \mathcal{A}(t)} \omega_x} \) \\
\bottomrule
\end{tabular}
\caption{Key market model definitions used in the financial market agent-based model algorithms: \ref{algo:lob state},  \ref{algo:fundamentalists}, \ref{algo:liquidtyprovider}, \ref{algo:chartists}, and  \ref{algo:injectsimulation} (see Figure \ref{fig:abm state flow}).} \label{tab:app:definitions}
\end{table}
\begin{algorithm}[H]
    \small
    \vspace{0pt}
    \SetAlgoLined
    \DontPrintSemicolon
    \KwIn{A confirmation or execution report string message from CoinTossX}
    \KwOut{The updated LOB state $\mathcal{L}(t)$}
    Extract order type, price, side, volume and trader identification fields\;
    \uIf{A new limit order has been confirmed}{
        \lIf{Buy limit order}{Add limit order $x_{t_i}$ to $\mathcal{B}(t)$}\lElse{Add limit order $x_{t_i}$ to $\mathcal{A}(t)$}
    }
    \uElseIf{A trade has been executed}{
        \lIf{Buy market order}{Remove the executed volume from the limit order having the corresponding ID in $\mathcal{A}(t)$}\lElse{Remove the executed volume from the limit order having the corresponding ID in $\mathcal{B}(t)$}
    }
    \ElseIf{A limit order has been cancelled}{
        \lIf{Buy limit order}{Remove the limit order having the corresponding ID from $\mathcal{B}(t)$}\lElse{Buy limit order}{Remove the limit order having the corresponding ID from $\mathcal{A}(t)$}
    }
    \eIf{$\mathcal{A}(t) \neq \emptyset$ and $\mathcal{B}(t) \neq \emptyset$}{
        Update $a(t)$ and $b(t)$ with the current updated LOB state $\mathcal{L}(t)$\;
    }{
        \lIf{$\mathcal{A}(t) \neq \emptyset$}{Update $a(t)$ with the current updated LOB state $\mathcal{L}(t)$}
        \lIf{$\mathcal{B}(t) \neq \emptyset$}{Update $b(t)$ with the current updated LOB state $\mathcal{L}(t)$}
    }
    Update $s(t)$ , $m(t)$ and $\rho(t)$ with the current or historic LOB state\;
    \caption{Algorithm for updating the state of the LOB with the latest execution message from CoinTossX \label{algo:lob state}}
\end{algorithm}
\newpage
\begin{algorithm}[H]
    \small
    \vspace{0pt}
    \SetAlgoLined
    \DontPrintSemicolon
    \KwIn{LOB state $\mathcal{L}_t$, agent parameters $f_t$ and simulation parameters $\bm{\theta}$}
    \KwOut{Market order $x_i = (0, \omega_{x_i}, t_{x_i})$}
    \lIf{$(m(t) - f_t) > \delta m(t)$}{$x_m = 50$}\lElse{$x_m = 20$}
    \eIf{$f_t < m(t)$}{
        $x_i$ is a sell order\;
        $\alpha = 1 - \frac{\rho(t)}{\nu}$
    }{
        $x_i$ is a buy order\;
        $\alpha = 1 + \frac{\rho(t)}{\nu}$\;
    }
    \lIf{The contra side of LOB $\mathcal{L}_t$ is non-empty}{Submit market order $x_i$ with volume $\omega_{x_i} \sim f(x; 10, \alpha)$ Power Law distribution with tail index parameter $\alpha$ (\cref{eq:1})}
    \caption{Fundamentalist agent action algorithm \label{algo:fundamentalists}}
\end{algorithm}

\begin{algorithm}[H]
    \small
    \vspace{0pt}
    \SetAlgoLined
    \DontPrintSemicolon
    \KwIn{LOB state $\mathcal{L}_t$ and simulation parameters $\bm{\theta}$}
    \KwOut{Limit order $x_i = (p_{x_i}, \omega_{x_i}, t_{x_i})$}
    Sample $u \sim U(0, 1)$\;
    \lIf{$u < (\rho(t) + 1) / 2$}{$x_i$ is a sell order}\lElse{$x_i$ is a buy order}
    \eIf{$x_i$ is a sell order}{
        $\alpha = 1 - (\rho(t) / \nu)$\;
        Sample $\eta \sim \Gamma \left(s(t), e^{\frac{\rho(t)}{\kappa}} \right)$\;
        Set the limit price as $p_{x_i} = b(t) + 1 - \lceil \eta \rceil$\;
        Set the volume as $\omega_{x_i} \sim$ Power Law distribution with tail index parameter $\alpha$\;
    }{
        $\alpha = 1 + (\rho(t) / \nu)$\;
        Sample $\eta \sim \Gamma \left(s(t), e^{-\frac{\rho(t)}{\kappa}} \right)$\;
        Set the limit price as $p_{x_i} = b(t) - 1 - \lceil \eta \rceil$\;
        Set the volume as $\omega_{x_i} \sim f(x; 10, \alpha)$ Power Law distribution with tail index parameter $\alpha$ (\cref{eq:1})\;
    }
    \caption{Liquidity provider agent action algorithm \label{algo:liquidtyprovider}}
\end{algorithm}

\begin{algorithm}[H]
    \small
    \vspace{0pt}
    \SetAlgoLined
    \DontPrintSemicolon
    \KwIn{LOB state $\mathcal{L}_t$, agent parameters $\tau$ and $\overline{p}_t$, and simulation parameters $\bm{\theta}$}
    \KwOut{Market order $x_i = (0, \omega_{x_i}, t_{x_i})$}
    Compute the agent's decision inter-arrival time $\Delta t$\;
    $\lambda = 1 - e^{- \Delta t / \tau}$\;
    Set the agent's EWMA $\overline{p}_t = \lambda (m(t) - \overline{p}_t)$\;
    \lIf{$(m(t) - \overline{p}_t) > \delta m(t)$}{$x_m = 50$}\lElse{$x_m = 20$}
    \eIf{$m(t) < \overline{p}_t$}{
        $x_i$ is a sell order\;
        $\alpha = 1 - \frac{\rho(t)}{\nu}$
    }{
        $x_i$ is a buy order\;
        $\alpha = 1 + \frac{\rho(t)}{\nu}$\;
    }
    \lIf{The contra side of LOB $\mathcal{L}_t$ is non-empty}{Submit market order $x_i$ with volume $\omega_{x_i} \sim f(x; x_m, \alpha)$ Power Law distribution with tail index parameter $\alpha$ (\cref{eq:1})}
    \caption{Chartist agent action algorithm \label{algo:chartists}}
\end{algorithm}

\begin{algorithm}[ht]
    \small
    \vspace{0pt}
    \SetAlgoLined
    \DontPrintSemicolon
    \KwIn{Agent and simulation parameters $\bm{\theta}$}
    Initialise agents with their submission times, order types and order IDs\;
    Set the initial LOB state $\mathcal{L}(0)$ with values for $s(0)$, $a(0)$, $b(0)$, $m(0)$\;
    Login a single client to CoinTossX to trade in a single security\;
    Open the market data feed socket by connecting to the correct port and IP address for the receival of UDP message packets\;
    Convert agent relative decision times to calendar time\;
    \ForEach{$x_i = (p_{x_i}, \omega_{x_i}, t_{x_i})$ with $i \in 1, \hdots, T$}{
        \uIf{$t_{x_i}$ is the liquidity providers' decision time}{
            \eIf{$x_i$ is a limit order}{
                Set order fields $p_{x_i}$ and $\omega_{x_i}$ according the liquidity provider decision rules\;
            }{
                Cancel the limit order having the current order's pre-assigned order ID\;
            }
        }
        \ElseIf{$t_{x_i}$ is the liquidity providers' decision time}{
            \If{The static price reference does not deviate from the bes by more than 10\%}{
                \uIf{Agent is a chartist}{
                    Set order fields $p_{x_i}$ and $\omega_{x_i}$ according the chartist decision rules\;
                }
                \ElseIf{Agent is a fundamentalist}{
                    Set order fields $p_{x_i}$ and $\omega_{x_i}$ according the fundamentalist decision rules\;
                }
            }
        }
        Receive UDP confirmation/execution message from CoinTossX that corresponds to the current order submitted\;
        Update the LOB state $\mathcal{L}(t)$\;
    }
    Disconnect from the market data feed port\;
    Log out from CoinTossX\;
    \caption{Basic logic for the submission of orders to CoinTossX \label{algo:injectsimulation}}
\end{algorithm}

\section{Expected stylised facts} \label{app:stylisedfacts}

We present a selection of stylised facts useful to constrain simple order book agent-based models. For more extensive lists of the empirical facts of financial markets we suggest readers refer to \citet{bouchaud2002statistical, chakraborti2011econophysics1, cont2001empirical, gould2013limit, lillo2004long, pagan1996econometrics}, and \citet{schmitt2017bimodality}. One key attribute of agent-based modelling is the idea that simple relational rules and interactions between agents can generate a variety of complex and dynamic in-sample distributional features; features that do not need to be directly (and distributionally) encoded into the model representation. Rather the distributional properties and various features emerge from the aggregate interaction between agents and their environment. 
\begin{table}[H]
    \centering
    \begin{tabular}{p{4.4cm}p{13cm}} 
    \toprule
    Feature & Description \\ 
    \midrule
  Return auto-correlations & The absence of auto-correlations in price fluctuations is sometimes cited as support for random walk models and the Efficient Market Hypothesis (EMH) because the presence of auto-correlations in the returns from financial markets would imply the existence of very simple strategies for making money using only knowledge of past share prices co-movements \cite{cont2001empirical}. Given the great incentives people have to make money, it follows that such opportunities would quickly be snatched upon by traders up until the point where the opportunity no longer exists through the cancelling out of auto-correlations by the behaviour of the traders. Thus, the existence of this stylised fact, when considering returns over a sufficient period of time is intuitive. More importantly as related to the order book, returns on bid-ask prices exhibit a significant negative first lag auto-correlation \cite{cont2001empirical}, suggesting fast mean reversion at the tick resolution. The absence of auto-correlations does not hold for time-scales greater than a week \cite{cont2001empirical}.  \\ 
  \midrule
  Heavy/Fat tails & \citet{mandelbrot1963variation}: The distribution of returns being leptokurtic, where a distribution is considered leptokurtic if its kurtosis is greater than that of the Normal distribution (kurtosis = 3).Returns at the extreme percentiles or tails are found to exhibit a power-law distribution with tail index $2 < \alpha < 5$ -- even after correcting returns for volatility clustering via GARCH-type models, the residual time series still exhibit fat tails. Suggested distributions include the generalized hyperbolic Student-t, the normal inverse Gaussian, the exponentially truncated stable, etc \cite{chakraborti2011econophysics1, pagan1996econometrics}. \\ \midrule
  Aggregational Gaussianity & \citet{kullmann1999characteristic}: The distribution of returns as the time-scale at which they are calculated $\Delta t$ increases, tends to normality. This feature becomes more apparent when calculating log-returns in trade time \cite{chakraborti2011econophysics1}. \\
  \midrule
  Volatility clustering & \citet{biais1995empirical} and \citet{mandelbrot1963variation}: The slow decay in the auto-correlations of the absolute value of price fluctuations is indicative of long memory, and large changes in financial markets returns tend to be grouped together. \\ \midrule
  Gain/Loss asymmetry & Large drops in prices tend to be observed more frequently than large price increases \cite{kahneman1979prospect}.\\ 
  \midrule
  Trade-sign auto-correlation & Order-flow in equity markets are persistent. That is, buy orders tend to be followed by more buy orders and sell orders tend to be followed by more sell orders. This feature is usually attributed to agent herding behaviour and trade splitting \cite{lillo2004long}. \\ 
  \midrule
  Depth profiles & Mean relative depth profiles exhibit a hump shape in a wide range of markets. The maximal mean depth available is most often reported to occur at $b_t$ and $a_t$ (at a relative price of 0) \cite{biais1995empirical, bouchaud2002statistical, gould2013limit}. \\ 
  \midrule
  Volume-volatility correlation & Trading volume is correlated with all measures of volatility. It has been shown that the variance of log-returns after \( N \) trades is proportional to $N$ \cite{plerou2000economic}. \\ 
  \midrule 
  Other desirable features & The \emph{bimodality of a security's price distortion} \cite{schmitt2017bimodality} --- a feature that is difficult to explain by the EMH, the \emph{Epps effect} \cite{CHANG2021126329,epps1979comovements}, the \emph{leverage effect} --- where most measures of volatility are correlated with returns, the \emph{power-law placement of limit orders around the best quotes}, and the \emph{Power-law distribution of trade inter-arrival times, order volumes, and limit order average lifetime} are other desirable features we would hope to be able to recover or conform to with a reasonably representative model of financial market interactions. \\ 
  \bottomrule
    \end{tabular}
    \caption{Summaries of some useful financial market stylised facts used to inform the model and parameter constraints.}
    \label{tab:orderbookfacts}
\end{table}

\section{Sensitivity analysis results \label{app:sensitivity analysis results}}

\begin{sidewaysfigure}
    \centering
    \includegraphics[width=.9\textwidth]{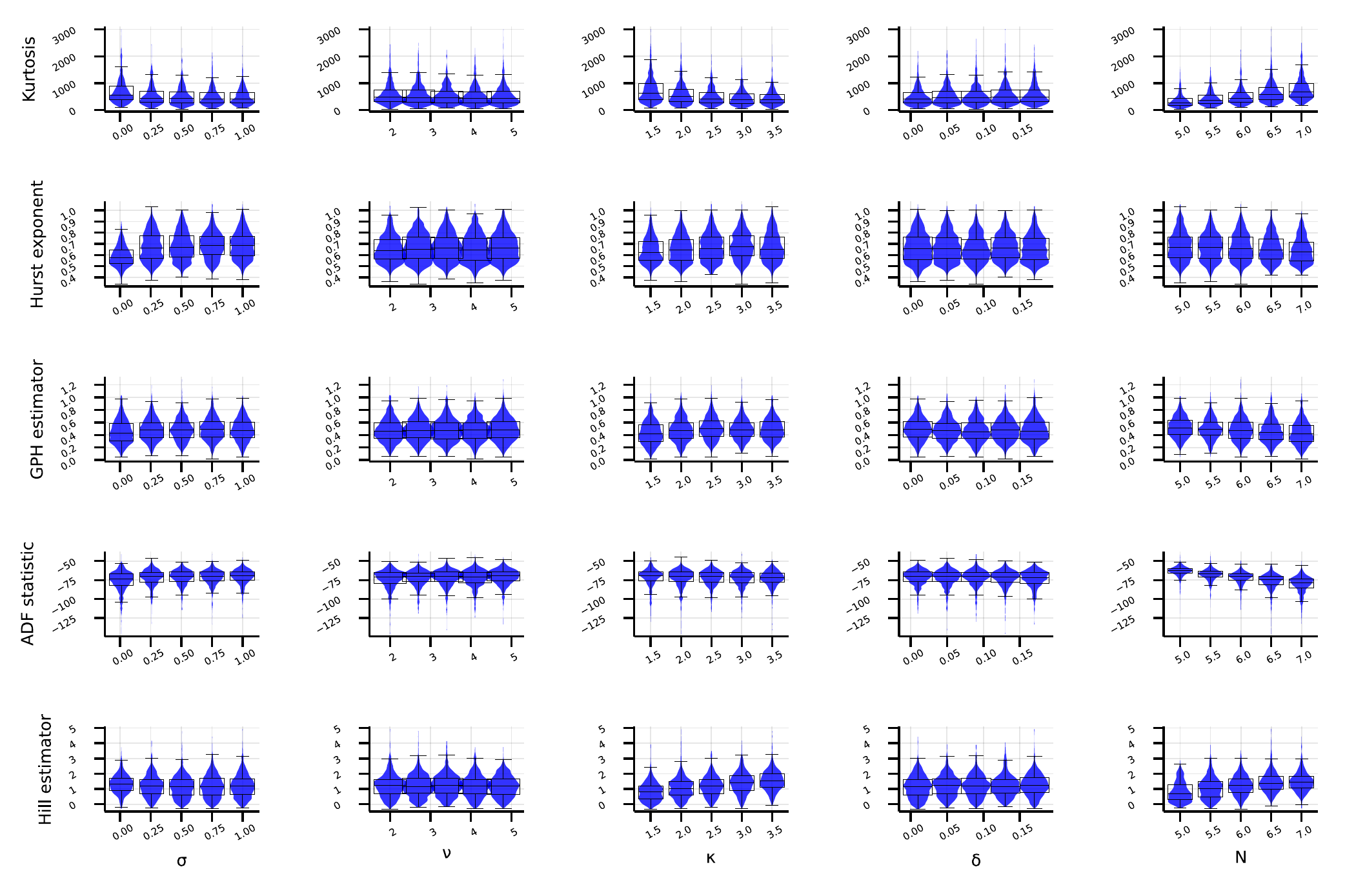}
        \caption{Mid-Price Moments: Summary statistics box-plots for each individual parameter-moment combination calculated on both simulated mid-price time-series. Columns represent individual moments while rows represent individual parameters. The vertical axis of each plot gives the values of the moments and the horizontal axis gives the unique parameter values chosen. \label{fig:sensitivity boxplot midprice}}
\end{sidewaysfigure}

\begin{sidewaysfigure}
    \includegraphics[width=.9\textwidth]{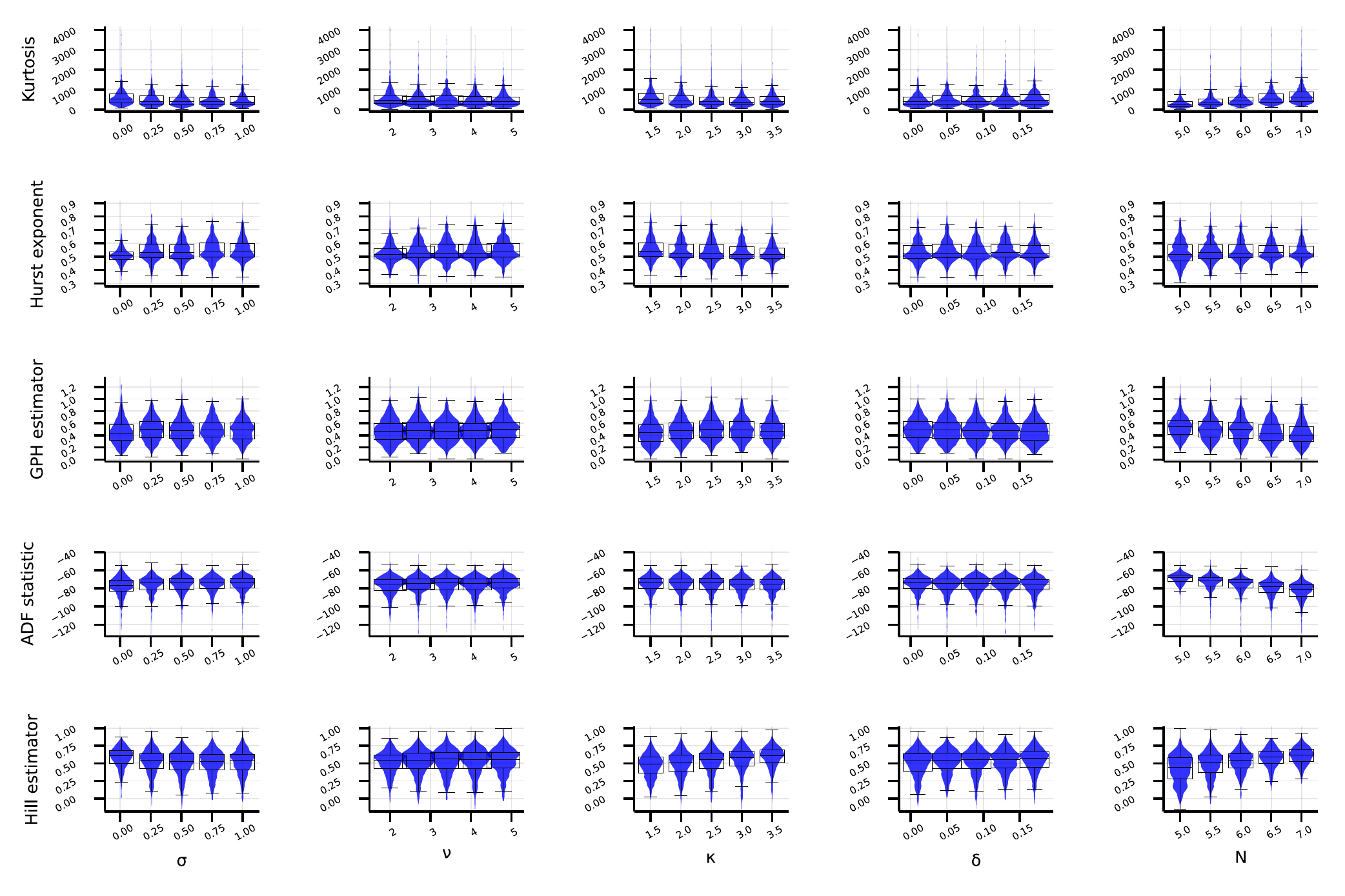}
    \caption{Micro-price Moments: Summary statistics box-plots for each individual parameter-moment combination calculated on both simulated micro-price. Columns represent individual moments while rows represent individual parameters. The vertical axis of each plot gives the values of the moments and the horizontal axis gives the unique parameter values chosen. \label{fig:sensitivity boxplot microprice}}
\end{sidewaysfigure}

\begin{figure}[p]
    \centering
    \subfloat[Mid-price Kurtosis surfaces for $\delta$ \& $\kappa$ and $N$ \& $\nu$]{\includegraphics[width=.45\textwidth]{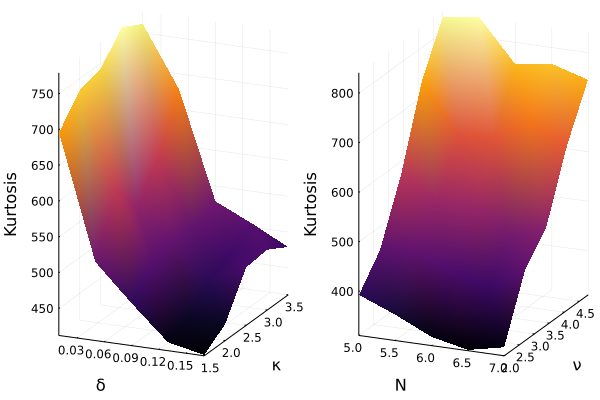}}
    \subfloat[Micro-price Kurtosis surfaces for $\delta$ \& $\kappa$ and $N$ \& $\nu$]{\includegraphics[width=.45\textwidth]{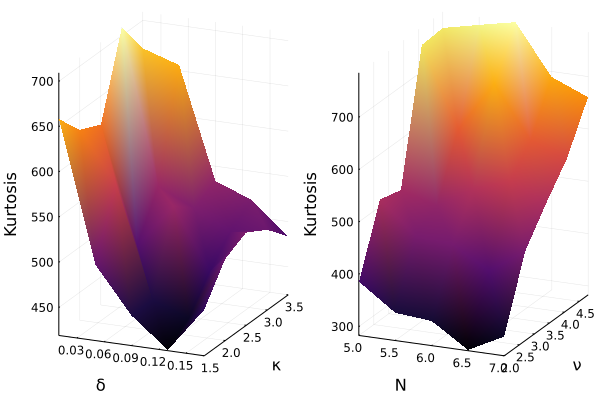}} \\
    \subfloat[Mid-price Hurst exponent surfaces for $\delta$ \& $\kappa$ and $N$ \& $\nu$]{\includegraphics[width=.45\textwidth]{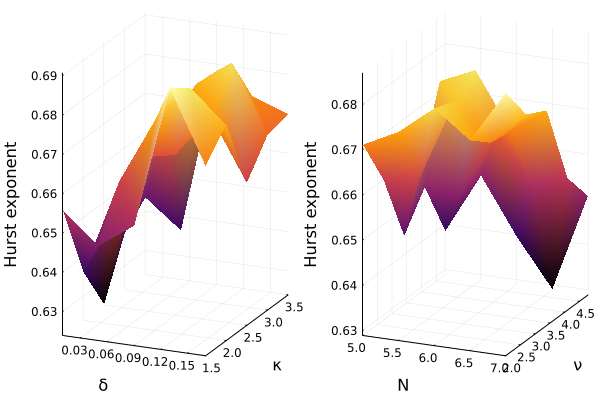}}
    \subfloat[Micro-price Hurst exponent surfaces for $\delta$ \& $\kappa$ and $N$ \& $\nu$]{\includegraphics[width=.45\textwidth]{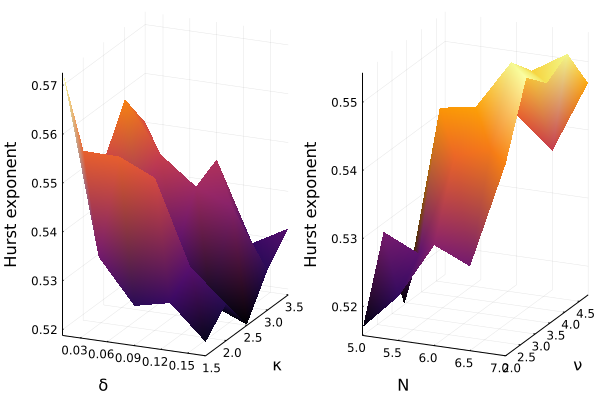}} \\
    \subfloat[Mid-price Hill estimator surfaces for $\delta$ \& $\kappa$ and $N$ \& $\nu$]{\includegraphics[width=.45\textwidth]{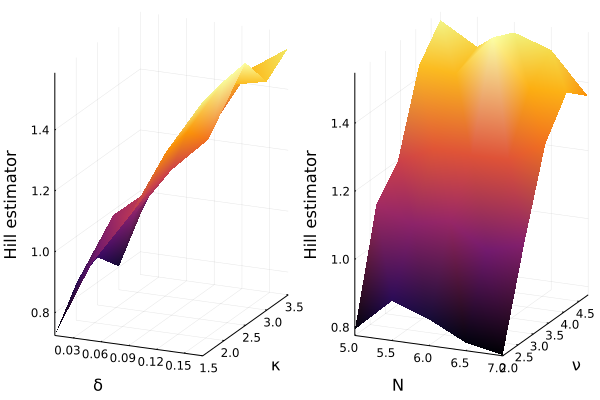}}
    \subfloat[Micro-price Hill estimator surfaces for $\delta$ \& $\kappa$ and $N$ \& $\nu$]{\includegraphics[width=.45\textwidth]{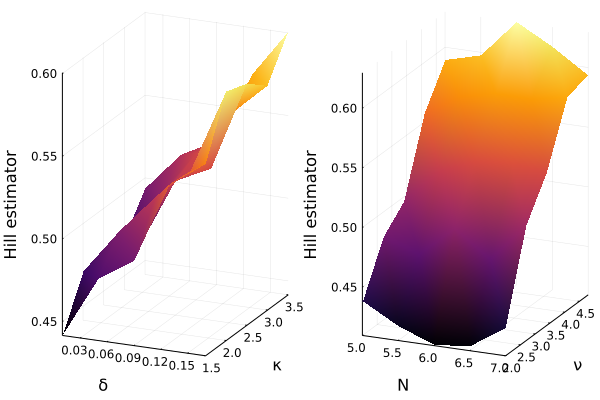}} \\
    \subfloat[Mid-price GPH statistic surfaces for $\delta$ \& $\kappa$ and $N$ \& $\nu$]{\includegraphics[width=.45\textwidth]{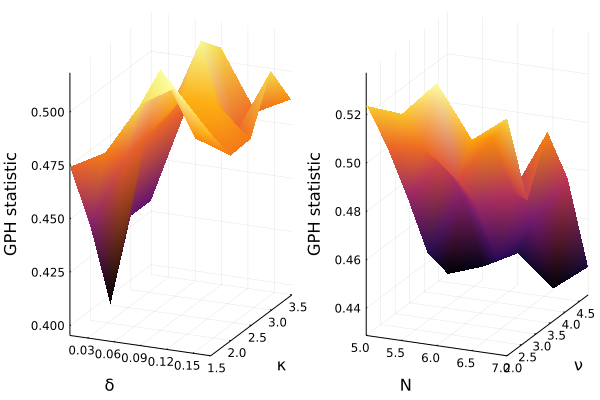}}
    \subfloat[Micro-price GPH statistic surfaces for $\delta$ \& $\kappa$ and $N$ \& $\nu$]{\includegraphics[width=.45\textwidth]{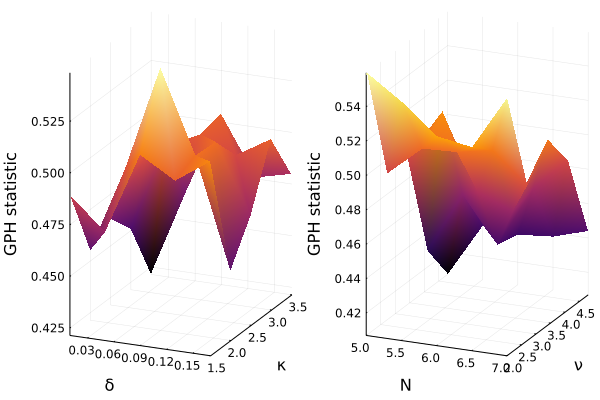}} \\
    \caption{Moment surfaces for pairwise combinations of parameters calculated on simulated log-returns and averaged over all combinations of other parameters \label{fig:moment surfaces}}
\end{figure}

\newpage
\section{Heuristic optimisation \label{app:nmta}}

\begin{figure}[!htb]
    \centering
    \subfloat[Convergence criterion and objective fitness values of the best vertex]{\includegraphics[width=.4\textwidth]{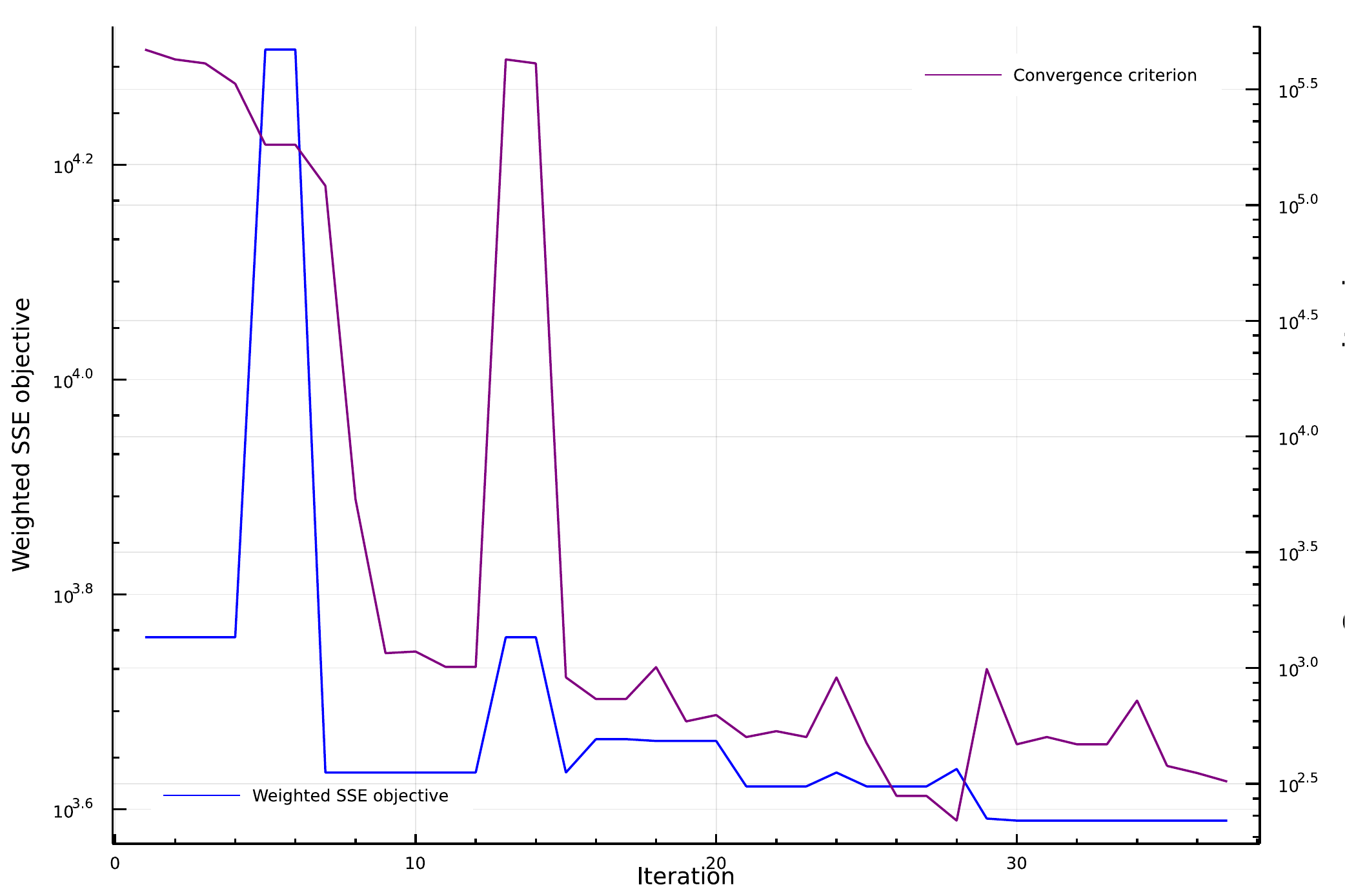}}
    \subfloat[Objective fitnesses of vertices in the simplex]{\includegraphics[width=.4\textwidth]{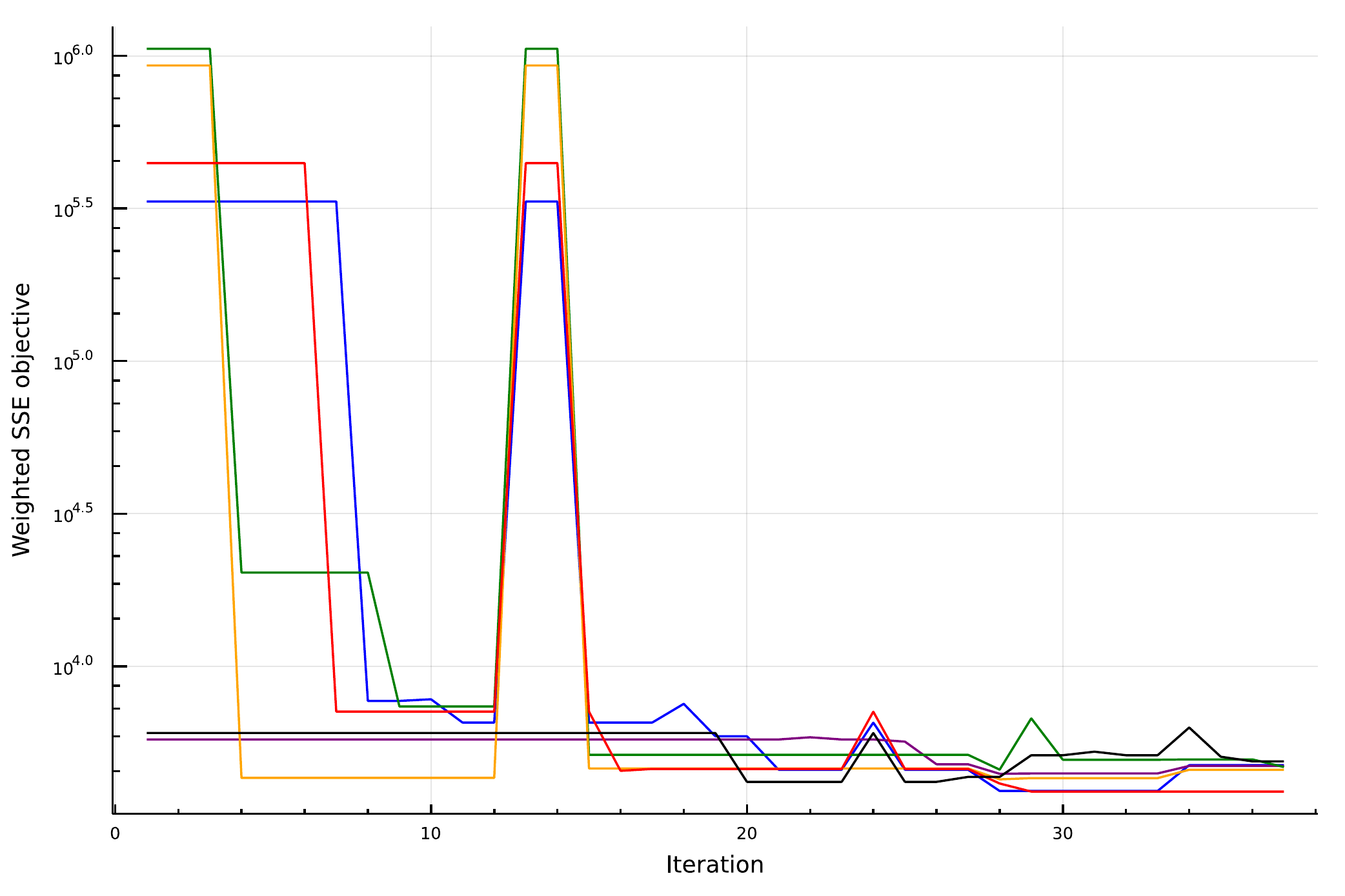}}
    \caption{NMTA optimization results are visualised through traces of the convergence criterion and simplex objective values \label{fig:nmta results}}
\end{figure}

Standard local optimisation methods are prone to failure because such an objective function does not always behave well to guarantee a global optimum solution. The objective function, built when taking into account the statistical properties of data, is not smooth or globally convex. There tends to be many local optima, so it is important to use an optimisation routine that is able to escape these local optima. Furthermore, due to the complexity of the problem and the nature of agent-based models, finding a global optimum is very computationally intensive, so any optimisation routine used should be able to efficiently explore the parameter space. Using the methodology proposed in the paper by \citet{gilli2003global}, we describe the Nelder-Mead with threshold accepting algorithm for the global optimisation of the objective function. In particular we follow the approach of \citet{gao2012implementing} when defining adaptive Nelder-Mead parameters that take into account the dimensionality of the optimisation problem.

In the Nelder-Mead (NM) simplex algorithm, each current solution has $n$ free moving parameters which consists of $n+1$ vertices of a simplex in the parameter space $\bm{\Theta}$. Each vertex represents a set of values for the free moving parameters of the model. At each iteration, the algorithm shifts the simplex according to either reflection, expansion, contraction and shrinkage. These steps are repeated until the iteration limit is reached or the convergence criterion is met. For non-convex functions such as the one dealt with here, the simplex shrinks rapidly to local minima.

The standard threshold accepting (TA) algorithm considers only a single point in the parameter space. A single step consists of shifting the value of a randomly chosen parameter to a neighbouring value (however that may be defined). If the new point obtained represents an improvement in fitness it is accepted as the new current solution. However, even if the new point is slightly worse, it will still be accepted as the new current solution provided the drop in fitness does not exceed a predefined threshold. This improves the ability of the algorithm to escape local minima in the objective function. The fitness thresholds $\bm{\tau} = \tau_1, \hdots, \tau_r$ are decreased at set length intervals/rounds to allow for convergence to a more globally optimal solution at the end of the optimisation. Four different relative thresholds are chosen to explore the local structure of the objective function along with a fixed number of steps $n_s = \{14, 12, 10, 8\}$ for each round respectively.

The Nelder-Mead simplex method provides an efficient way of identifying the search direction while threshold accepting helps to avoid local minima. Now considering a combination of the two algorithms, each iteration uses either the Nelder-Mead search or Threshold Accepting random shift with a probability of $(1 - \xi)$ and $\xi$ respectively. The Nelder-Mead search will have a higher probability of being chosen. In the TA algorithm, the magnitude of the random shift is dependent on the mean value (over the simplex) of the parameter being shifted. In addition to the TA algorithm, threshold accepting is also applied to the simplex search by ensuring that reflection, expansion, contraction and shrinkage perturbations are evaluated against threshold adjusted fitness such that a worse fitness may be accepted in the simplex search as well.

The trace results of the optimisation routine are visualised in Figure \ref{fig:nmta results}. Close to convergence was reached near 40 iterations with no thresholds being applied in the last few iterations. Vertices in the simplex were close to converging to the best with no significant improvement in the best vertex for subsequent iterations.\footnote{Optimisation with CoinTossX and the Julia-Java wrapper was initially prone to crashes in the Julia environment. It was suspected to be a Julia memory and garbage collection issue that came with very long consecutive simulations. It is noted that CoinTossX was designed purely as a matching engine for simulations, rather than a tool to be used for calibration.}

\end{document}